\documentclass[a4paper,12pt]{article}
\usepackage{natbib}
\usepackage{hyperref}
\usepackage{epsf}
\usepackage{epsfig}
\usepackage{multirow}
\usepackage[english]{babel}
\usepackage{graphicx}
\usepackage{geometry}
\usepackage{subfigure}
\usepackage{natbib}
\usepackage{amssymb}
\usepackage{amsmath}
\usepackage{color}
\usepackage{enumerate}
\usepackage{authblk}

\newcommand{\blambda}{\mbox{\boldmath{$\lambda$}}}
\newcommand{\bepsilon}{\mbox{\boldmath{$\epsilon$}}}
\newcommand{\bmu}{\mbox{\boldmath{$\mu$}}}
\newcommand{\bxi}{\mbox{\boldmath{$\xi$}}}
\newcommand{\bfeta}{\mbox{\boldmath{$\eta$}}}

\title{\textbf{A dynamic probabilistic
principal components model for the analysis of longitudinal metabolomics data.}}

\begin{document}
\pagestyle{plain}

\author[1]{Gift Nyamundanda}
\author[1]{Isobel Claire Gormley\thanks{claire.gormley@ucd.ie}}
\author[2]{Lorraine Brennan}
\affil[1]{{\footnotesize School of Mathematical Sciences, University College Dublin, Ireland.}}
\affil[2]{{\footnotesize School of Agriculture and Food Science, Conway Institute, University
College Dublin, Ireland.}}

\maketitle

\begin{abstract}
In a longitudinal metabolomics study, multiple metabolites are measured from several observations at many time points. Interest lies in reducing the dimensionality of such data and in highlighting influential metabolites which change over time. A dynamic probabilistic principal components analysis (DPPCA) model is proposed to achieve dimension reduction while appropriately modelling the correlation due to repeated measurements. This is achieved by assuming an autoregressive model for some of the model parameters. Linear mixed models are subsequently used to identify influential metabolites which change over time. The proposed model is used to analyse data from a longitudinal metabolomics animal study.
\end{abstract}

\section{Introduction}
\label{sec:intro}

Metabolomics is the study of low molecular weight compounds
known as metabolites found in biological samples; its application
reveals information on metabolic pathways within an organism. The
number of areas in which metabolomics is applied has recently
enjoyed rapid growth and metabolomics is now employed in
fields such as nutrition, toxicology and disease diagnosis. In a
typical metabolomics study large data sets are generated using
analytical technologies such as nuclear magnetic resonance spectroscopy (NMR)
\citep{Reo2002} and mass spectrometry (MS) \citep{Dettmer07}. With respect to
NMR spectroscopy the resulting spectrum consists of a series of peaks where the
height of a peak is related to the relative abundance of the associated
metabolite. Studying such metabolomic profiles gives insight to the metabolic
state of a system.\\

Metabolomic data sets are usually high-dimensional, in that the
resulting spectra contain many peaks (i.e. variables), yet they are
characterised by small sample sizes -- hence classical statistical
approaches cannot be easily applied. The data sets contain
variables that are not independent in that metabolites can be represented by
more than one peak and metabolites can be highly correlated
\citep{vandenBerg2006}. In
addition to correlated variables, in longitudinal metabolomics data
sets there is further correlation structure due to the repeated
measurements of observations over time. Hence, appropriate statistical
models are required in order to appropriately model the data and extract true, important information.\\

Within the metabolomics literature, principal components analysis
(PCA) \citep{Jolliffe02} is often used for multivariate data
exploration \citep{Walsh07, smolinska2012, cassol2013, carvalho2013, bathen2013, sachse2012}. Methods that improve and extend the
application of this common statistical technique will prove
extremely useful to the metabolomics practitioner, and to scientists
in other fields. The application of PCA to longitudinal studies is limited however by
the fact that PCA does not take into account information about the
experimental design i.e. if PCA is applied to all time points
simultaneously, measurements taken repeatedly
over time are assumed independent \citep{Choi06}. In such a case, since PCA
looks for directions in the data space with maximum variation, time
related variation will act as a confounding factor obscuring
potential differences due to
treatment.\\

Several extensions to PCA have been developed to take into account
the experimental design of a study and therefore can be used to
analyse longitudinal metabolomics data more appropriately. These
include weighted PCA \citep{Jansen04} which uses weights to account
for variation due to repeated measurements and ASCA \citep{smilde05}
which combines analysis of variance and simultaneous components
analysis methods to deal with complex multivariate datasets.
\cite{Jansen09} employ local PCA models at each time point, and then
link these local models to each other. Dynamic PCA \citep{Smilde10}
uses a back-shift matrix to analyse data from multiple time points
simultaneously. The main limitation of these approaches is that they
do not have an associated generative probabilistic model. Hence, it
is difficult to assess the uncertainty in the fitted model estimates,
and model extensions are not feasible.\\

Mixed effects models have also been employed to model longitudinal
metabolomics data. \cite{Mei09} employ a linear mixed-effects model
(LMM) in the context of feature selection for longitudinal
metabolomics data, but under the assumption that spectral peaks are
independent variables. The high levels of correlation between
spectral peaks (i.e. metabolites) is biologically important however,
and such correlation structure should be explicitly modeled. In a
similar vein, \cite{Berk11} employ smoothing splines mixed-effects
models to model longitudinal metabolomics data. While these models
have a statistical modelling basis and therefore appropriately model
the longitudinal aspect of the data, multiple testing issues
\citep{Dudoit03} result as the chances of false positives increase
with the dimensionality of the data. While this problem can be
controlled \citep{Benjamini95}, dimension reducing
features of methods such as PCA are attractive.\\

Probabilistic PCA (PPCA) is an approach to PCA based
on a Gaussian latent variable model \citep{tipping99, nyamundanda10}. 
PPCA retains the benefits of PCA, such as dimension reduction, while facilitating
model extensions through its basis in a statistical model. Here an
extension of PPCA called dynamic PPCA (DPPCA) is proposed which
allows PPCA to appropriately model the time dependencies in
longitudinal metabolomics data. This is achieved by assuming a stochastic
volatility model for some of the PPCA parameters. The
proposed DPPCA model is closely related to the dynamic factor
analysis model
\citep{aguilar00} employed to model multivariate financial time series data.\\

Data generated in longitudinal metabolomics studies form the basis for the
development of the proposed DPPCA model. Examples of such studies include, but
are not limited to, postprandial human studies and long term drug treatment
studies  \citep{Wopereis09, Lin11, Krug12, Nicholson12}. Interest lies in
reducing the dimensionality of the
data (for statistical and visualisation purposes) and subsequently highlighting
influential metabolites which
change over time, while appropriately modelling the longitudinal nature of the
data. The proposed DPPCA model is
 employed to achieve dimension reduction and model the time dependencies; linear
mixed models (LMM) are then
 employed to identify the metabolites which change over time. The utility
of the DPPCA approach is demonstrated through the analysis of data from a
longitudinal metabolomics animal study.\\

The remainder of the article is structured as follows. An overview of
longitudinal metabolomics studies is presented in Section \ref{sec:data}. The
DPPCA model is introduced in
Section \ref{sec:DPPCA} and the use of stochastic volatility models to account
for
the correlation due to repeated measurements is detailed. The
DPPCA model is estimated within the Bayesian paradigm; accordingly
Section \ref{sec:modelfitting} specifies the necessary prior distributions and
describes
the use of Markov chain Monte Carlo (MCMC) techniques to fit the
DPPCA model. Section \ref{sec:results} details the application of the DPPCA
model to
a longitudinal metabolomics data set. Discussion of the developed model and
further avenues of research are deferred until the conclusion, in Section
\ref{sec:discussion}.

\section{Longitudinal metabolomics studies}
\label{sec:data}

In recent years, a number of longitudinal metabolomics datasets have emerged in
the literature \citep{Wopereis09, Lin11, Krug12}. With regard to human
applications, a number of studies employing metabolomics over time following
acute challenges such as the oral glucose tolerance test have recently been
published and shown to be extremely powerful in studying subtle changes.
Applying metabolomics to longitudinal animal studies for determining long term
drug toxicity and efficacy is also an important emergent area. In such
applications a number of key study aims typically exist which, in general, can be
described as follows:
 
\begin{enumerate}
\item[(i)] data visualisation
\item[(ii)] assessing the effect of time within each treatment group and
\item[(iii)] identifying metabolites which change over time within each
treatment group.
\end{enumerate}

The DPPCA model proposed here helps address these specific aims. In the case of (i) the DPPCA model facilitates visualisation of the study participants in a reduced dimensional space, while appropriately modelling the time course nature of the data. The effect of time within each treatment group (aim (ii)) can be assessed by applying the DPPCA model to the data from each treatment group. An additional output of the DPPCA model is a list of the most influential metabolites within each group. To address aim (iii) univariate analyses with LMM are then carried out to identify those influential metabolites which change over time.\\

Metabolomics data from a longitudinal animal study motivate and illustrate the
proposed DPPCA model. The study has been described in detail in
\cite{Carmody10}. Briefly, an animal model of epilepsy was employed by repeated
administration of pentylenetetrazole (PTZ) which leads to the development of
generalised tonic-clonic seizures.  Over the administration period (5 weeks)
urine samples were collected from treated animals (PTZ treated) and control
animals (saline treated animals). The aim of the study was to determine
metabolic changes that occur over time during PTZ treatment. \\

\begin{figure}[h!]
\begin{center}
\includegraphics[width=0.95\textwidth,height=0.4\textheight]{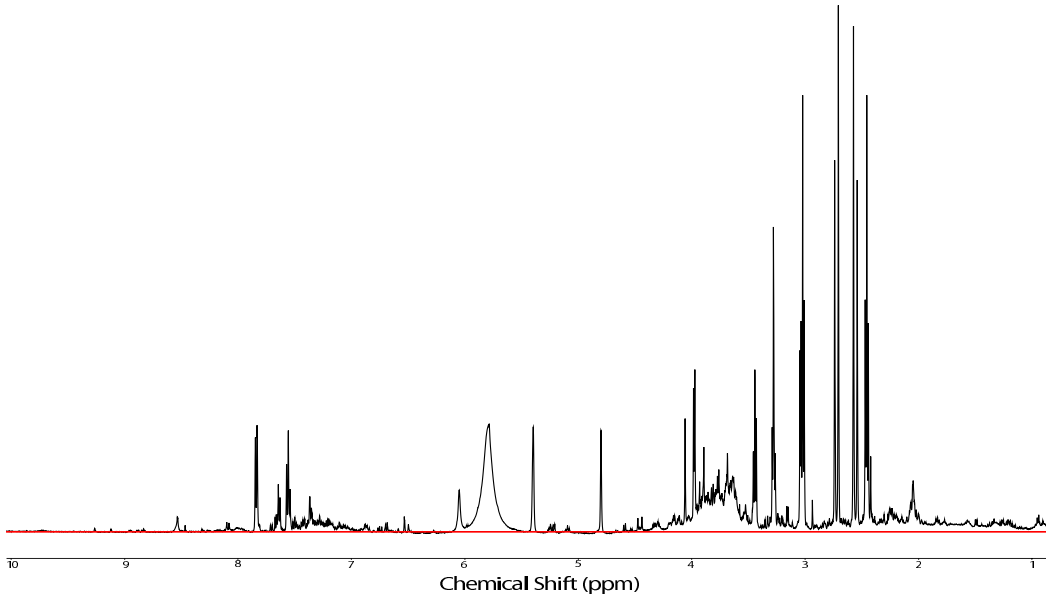}
\caption{A metabolomic profile resulting from the urine sample collected at a
single time point from an animal in the longitudinal metabolomic study.}
\label{fig:samplespectra}
\end{center}
\end{figure}

NMR spectra were acquired from the urine samples and the spectra were integrated
into bin regions of 0.04 parts per million (ppm), excluding the water regions
(4.0--6.0 ppm). For the purposes of this work, the final acquired data set
consists of NMR spectra for $n = 15$ animals (8 treated and 7 control), each
containing $p = 189$
spectral bin regions, from $M = 8$ time points. The $p = 189$ peaks in the
spectra at different chemical shift values (measured in ppm) relate to specific
metabolites; the height of a peak in any spectrum details the relative abundance
of the associated metabolite in the animal's urine sample. Figure
\ref{fig:samplespectra} illustrates a metabolomic spectrum resulting from the
urine sample collected at a single time point from an animal in the study. \\

\section{Dynamic Probabilistic Principal Components Analysis}
\label{sec:DPPCA}

Probabilistic principal components analysis (PPCA) is a latent factor model constrained such that the maximum likelihood estimates of the parameters span the principal subspace of conventional PCA. Given its underlying assumptions however, PPCA is only applicable to data from a
cross sectional study. Here an extension of PPCA to a dynamic PPCA (DPPCA) model
is developed; a brief introduction to PPCA, and its extension to the DPPCA model, are detailed
in what follows.

\subsection{Probabilistic Principal Components Analysis (PPCA)}

PPCA is a generative statistical model which models a high-dimensional observed
data point as a linear function of a corresponding low-dimensional latent
variable plus isotropic (full-dimensional) noise. For each of $n$ animals, let $\mathbf{x}_{i}^{T}=(x_{i1}, \ldots, x_{ip})$ denote the set of $p$ observed variables for animal $i$ (eg. an NMR
spectrum with $p$ spectral bins). The PPCA model relates each $\mathbf{x}_{i}$
to a $q$-dimensional latent Gaussian variable $\mathbf{u}_{i}$ (typically $q \ll p$) through the
linear model:
$$\mathbf{x}_{i} = \emph{W} \mathbf{u}_{i} + \bepsilon_{i}$$
where $\emph{W}$ is a $p \times q$  loadings matrix and the error term  $\bepsilon_i$ is
assumed to have a multivariate Gaussian distribution, centred at zero
with covariance $\sigma^2\emph{I}$, where \emph{I} denotes the identity
matrix. The error term models the part of the observed data which cannot be accounted for by the $q$ underlying latent
variables, or principle components (PCs). Assuming a standard multivariate normal (MVN) distribution for $\mathbf{u}_i$, each
data point has a zero mean multivariate normal distribution with covariance $\emph{W}\emph{W}^{T}
+ \sigma^2\emph{I}.$

Crucially, the likelihood of the PPCA model is maximized when the columns of
$\emph{W}$ span the principal subspace of conventional PCA \citep{tipping99}.
Thus the maximum likelihood estimate of the loadings matrix in PPCA corresponds
exactly to the loadings matrix in conventional PCA. Hence the model output in PPCA is exactly that obtained in
conventional PCA, but with the additional advantages of uncertainty assessment
and potential model extensions.

\subsection{Dynamic Probabilistic Principal Components Analysis (DPPCA)}
\label{sec:dppca2}

The derivation of PCA from a probabilistic framework facilitates the development
of dynamic PPCA as a tool for modelling longitudinal multivariate data. Under the
DPPCA model, the set of $p$ observed variables $\mathbf{x}_{im}$ for animal
$i$ at time point $m$ ($m = 1, \ldots, M$) is modeled as:
\begin{eqnarray}
\mathbf{x}_{im} & = & \emph{W}_m \mathbf{u}_{im} +
\bepsilon_{im}
\label{eqn:DPPCAeqn}
\end{eqnarray}
where $\emph{W}_m$, the loadings, and $\mathbf{u}_{im}^{T} = (u_{i1m},
\ldots, u_{iqm})$, the latent scores, vary with time. \\

Unlike the PPCA model which constrains the covariance of the multivariate
Gaussian distribution of the latent variables to be an identity matrix, the
DPPCA model eases the equal variance restriction such that
\begin{eqnarray*}
p(\mathbf{u}_{im}) =  \mbox{MVN}_{q}(\mathbf{0}, \emph{H}_m)
\label{eqn:scoresdist}
\end{eqnarray*}
where $\emph{H}_m = \mbox{diag}(h_{1m}, \ldots, h_{qm})$. This assumption
allows the variances of the underlying latent variables to differ across the
latent dimensions and to depend on time.\\

The error, $\bepsilon_{im}$, for animal $i$ at time $m$ is also
assumed to have a multivariate Gaussian distribution:
$$p(\bepsilon_{im}) = \mbox{MVN}_{p}(\mathbf{0},
\sigma_m^2\emph{I}).$$
Again, the variance parameter $\sigma_{m}^{2}$ varies with time. The errors,
$\bepsilon_{im}$ and the latent variables (or scores),
$\mathbf{u}_{im}$ are assumed to be mutually independent for all $m = 1,
\ldots, M$.\\

While the variance parameter of the error terms $\sigma_{m}^{2}$ varies with
time, it is constrained to be constant across all observed variables. This is in
line with the assumptions of the underlying PPCA model; should the variances be
unconstrained across variables a dynamic factor analytic model results \citep{mcnic08, aguilar00}.
Thus the DPPCA model can be viewed as a constrained dynamic factor model.\\

The choice of developing the DPPCA model, rather than employing an alternative dynamic factor model to analyse the metabolomic data under study, deserves explanation. The manner in which time dependence is accounted for in the DPPCA model, and the constraints employed, are motivated by the explicit needs of the motivating metabolomics application. The metabolomics practitioners are interested in time evolving metabolites, hence the need for a different loadings matrix at each time point, leading to a highly parameterised model. Further, strongly motivated by the ubiquitous use, understanding and acceptance of PCA in the metabolomics field \citep{smolinska2012, cassol2013, carvalho2013, bathen2013, sachse2012}, maintaining a link to PPCA was deemed to be highly desirable. As the link to PPCA occurs by constraining the error variances to be equal, this modelling decision satisfied the metabolomic scientists, 
and provided a more parsimonious model than a generic dynamic factor model. The appropriateness of the DPPCA model assumptions are assessed after model fitting in Section \ref{sec:modelfit}, using posterior predictive model checking.

\subsection{Stochastic Volatility Models}

Stochastic volatility models \citep{Jacquier94, kim&shephard98} are popular in
econometrics and finance where they are typically employed to model the variance
of returns over time, which are highly correlated. The DPPCA model accounts for
the correlation due to repeated measurements through the use of stochastic
volatility (SV) models. Specifically, the DPPCA model assumes that at time point
$m$ the variances $h_{1m}, \ldots, h_{qm}$ of the latent variables and the error
variances $\sigma_m^2$ follow a latent stochastic process. These assumptions
allow the DPPCA model to account for any potential time dependence in
longitudinal multivariate data.\\

Again, the motivation behind the incorporation of SV models in DPPCA requires explanation. While SV models typically model settings with many time points \citep{aguilar00}, they have been employed when modelling longitudinal multivariate data, where the number of time points is low. \cite{Ramoni2002}, \cite{Wu2005} and \cite{Wang2008}, for example, employ SV models for modelling high dimensional time course data where the number of time points ranges from 8 to 18. Hence the SV model was deemed suitable to model the evolution of the latent variables over time. The appropriateness of the SV model assumptions is assessed after model fitting in Section \ref{sec:modelfit}.

\subsubsection{A stochastic volatility model for the latent variables}

An SV model on the latent variable $u_{ijm}$ of animal $i$ ($i = 1,
\ldots, n$) for principal component $j$ ($j = 1, \ldots, q$) at time
point $m$ ($m = 1, \ldots, M$) can be expressed as:
\begin{eqnarray*}
u_{ijm} & = & \exp(\lambda_{jm}/2)\zeta_{ijm}
\end{eqnarray*}
where $\lambda_{jm} = \log(h_{jm})$ is known as the log volatility
and $\zeta_{ijm}$, which  has a standard univariate Gaussian
distribution, denotes the error term of the SV model. Thus the
conditional distribution of the latent variable is
$u_{ijm}|\lambda_{jm} \sim N[0, \exp(\lambda_{jm})]$. The $q$-vector
of log volatilities, $\blambda_{m}^{T} = (\lambda_{1m},
\ldots, \lambda_{qm})$, is assumed to have a stationary first order
vector autoregressive process VAR(1) centered around a mean
$\bmu^{T} = (\mu_1, \ldots, \mu_q)$:
\begin{eqnarray*}
\blambda_m &=& \bmu + \it{\Phi}(\blambda_{m-1} - \bmu) + \mathbf{R}_{m}
\end{eqnarray*}
where $\it{\Phi}$ is a matrix of persistence parameters and $\mathbf{R}_m
\sim \mbox{MVN}_{q}(\mathbf{0}, \it{V})$ are independent innovations. The
model restricts dependencies across the principal dimensions by constraining the
matrix of persistence parameters $\it{\Phi}$ and the covariance of the
innovations $\it{V}$ to be diagonal i.e. $\it{\Phi} =
\mbox{diag}(\phi_1, \ldots, \phi_q)$ and $\it{V} = \mbox{diag}(v_1^2,
\ldots, v_q^2)$ respectively. The innovation variance $v_j^2$ is the uncertainty
associated with predicting the current log volatility using the log volatility
from the previous time point on component $j$.  The persistence parameter
$\it{\Phi}$ is the parameter of interest; it measures the strength of the
relationship between time points. For stationarity, the persistence parameter
$\phi_j$ is constrained to lie between -1 and 1 \citep{kim&shephard98}. The
initial state, by stationarity, is drawn from the model
$p(\blambda_{1}) = MVN_{q}[\bmu,\mbox{diag}(\frac{v_1^2}{1-\phi_1^2}, \ldots,
\frac{v_q^2}{1-\phi_q^2})]$. The distribution of the log volatilities
$\blambda_{m}$ given the log volatilities of the previous time point
$\blambda_{m-1}$ is given by $\mbox{MVN}_{q}[\bmu +
\it{\Phi}(\blambda_{m-1} - \bmu),\it{V}]$ for $m >
1$.\\

Constraining the covariance matrix $\it{V}$ to be diagonal is a modelling decision motivated by the fact that the PPCA model does not facilitate dependence across the principal components and PPCA underpins the DPPCA model, as detailed in Section \ref{sec:dppca2}. Such a model was considered by \cite{harvey1994}, \cite{kim&shephard98} and \cite{jacquier1995} among others; \cite{aguilar00} allow correlation across dimensions, motivated by their financial application area.

\subsubsection{A stochastic volatility model for the errors}
Additionally, another SV model is adopted to model the potential
time dependence in the errors of the DPPCA model. The $p$-vector of
errors of observation $i$ at time $m$ can be expressed as
$\bepsilon_{im} = \exp[\eta_m/2]\bxi_{im}$
where $\eta_m = \log(\sigma_m^2)$ is the log volatility at time $m$ and
$\bxi_{im} \sim
MVN_{p}(\mathbf{0}, \emph{I})$. The log volatilities $\eta_m$
on the errors are assumed to have a stationary
first order autoregressive process AR(1):

\begin{eqnarray*}
\eta_m = \nu + \phi (\eta_{m-1} - \nu) + r_{m}
\end{eqnarray*}
where the center of the AR(1) model is $\nu$ and the persistence
parameter $\phi$ is constrained such that $\phi \in [-1,1]$.
The innovations of the AR(1) model are assumed to be
normally distributed, $r_{m} \sim  N(0,v^2)$. It follows that the
initial state of the SV model is $p(\eta_1) =
N(\nu, \frac{v^2}{1-\phi^2})$ and that $p(\eta_m|\eta_{m-1}) =
N[\nu + \phi(\eta_{m-1} - \nu),v^2]$ for $m > 1$. Note that, as stated in Section \ref{sec:dppca2}, to maintain the link to PPCA and for
reasons of parsimony, each of the $p$ dimensions in the error
$\bepsilon_{im}$ are constrained to follow the same AR(1)
model.

\section{Estimation of the DPPCA model}
\label{sec:modelfitting}

Under the DPPCA model, the full augmented data likelihood function based on the
data $\emph{X} = (\emph{X}_{1}, \ldots, \emph{X}_{n})$ and the latent variables $\emph{U} = (\emph{U}_{1}, \ldots \emph{U}_{n})$, $\it{\Lambda} = (\blambda_{1}, \ldots, \blambda_{M})$ is:

\begin{eqnarray*}
p(\emph{X}, \emph{U}, \it{\Lambda}, \bfeta | \emph{W},
\theta_1,\theta_2) & = & \left[ \prod_{m=1}^M \prod_{i=1}^n p(\mathbf{x}_{im}
| \emph{W}_m, \mathbf{u}_{im}, \eta_m) p(\mathbf{u}_{im} |
\blambda_m) \right] p(\bfeta | \theta_1) p(\it{\Lambda}
|\theta_2)
\end{eqnarray*}
where $\theta_1 = (\nu, \phi, v^{2})$ and $\theta_2 = (\bmu,
\it{\Phi}, \it{V})$ denote the SV model parameters on the errors and latent scores respectively. The PPCA model on each time point
$p(\mathbf{x}_{im}|\emph{W}_m, \mathbf{u}_{im}, \eta_m)$ is
MVN$_{p}[\emph{W} \mathbf{u}_{im}, \exp(\eta_{m})\emph{I}]$.\\

A Bayesian approach is taken when estimating the DPPCA model; this requires the
specification of prior distributions for all the model parameters. The resulting
posterior distribution is intricate and Markov chain Monte Carlo methods are
necessary to produce realizations of the model parameters. Specifically, a
Metropolis-within-Gibbs algorithm is required to sample from the full
conditional distributions for all model parameters and latent variables.\\

\subsection{Prior distributions}

Prior distributions over the full set of the model parameters need to be
specified.  It is assumed that the prior distributions on the model parameters
are independent. Under the PPCA part of the DPPCA model, the only parameters are
the loadings matrices $\emph{W}_{1}, \ldots, \emph{W}_{M}$. A
$q$-dimensional multivariate normal prior distribution, centered at
$\mathbf{0}$ with covariance $\it{\it{\Omega}}_m$, is assumed for each row of the
loadings matrix $\emph{W}_{m}$ at time $m$.  \\

The remaining model parameters are all parameters of the SV part of the DPPCA
model. Non-informative normal prior distributions are specified on the means of
the SV models i.e. a $N(\mu_{\nu}, \sigma_\nu^2)$ distribution is specified for
$\nu$ and a $N(\mu_{\mu}, \sigma_\mu^2)$ distribution is assumed on each of the
univariate elements of $\bmu$, where the variance hyperparameter in
each of these priors is large. A conjugate prior is assumed for the variances of
the innovations in the SV models i.e. an inverse gamma $IG(\alpha/2,\beta/2)$
distribution is chosen for the prior distribution of $v^2$ and for each of the
diagonal elements of $\it{V}$. For stationarity, the persistence parameters
of the SV models are constrained to lie in $[-1, 1]$; accordingly the prior
distributions on $\phi$ and on the diagonal elements of $\it{\Phi}$ are
truncated normal distributions, $N_{[-1,1]}(\mu_\phi, \sigma_\phi^2)$.\\

As in any Bayesian setting, the choice of prior distribution can potentially
influence parameter inference. Sensitivity analyses were conducted to assess the
influence of different choices of priors on the resulting posterior
distribution. Some sensitivity was observed in the case of the persistence
parameters. \cite{kim&shephard98} employ a transformed beta prior for the persistence parameters, but sensitivity analyses here suggested that the posterior distribution strongly depended on the values of the hyperparameters used. In a similar setting to the DPPCA model, \cite{aguilar00} employ a truncated (between $\pm 1$) Gaussian prior for the persistence parameters; the posterior distributions were less sensitive to the parameter specification under this prior. Thus, a Gaussian prior, truncated (between $\pm 1$), was employed here for the persistence parameters.

\subsection{The Metropolis-within-Gibbs sampler}
\label{sec:MwG}

Given the specified prior distributions, the resulting posterior distribution
 is intricate and Markov chain Monte Carlo (MCMC) methods are
required to produce realizations of the model parameters. The full conditional
distributions for the loadings matrices $\emph{W}_m$, the latent scores
$\emph{U}_{m}$, the SV model means $\nu$ and $\bmu$, and the SV
model innovation variances $v^2$ and $\it{V}$ exist in standard form,  and a
straightforward Gibbs sampler can be employed to draw samples. However, the
full conditional distributions for the persistence parameters $\phi$ and
$\it{\Phi}$ and for the log volatilities $\it{\Lambda}$ and $\bfeta$
are not available in closed form; values from these distributions are therefore
sampled using a Metropolis Hastings step. Hence a Metropolis-within-Gibbs
algorithm \citep{Gilks96}  is required to sample from the full conditional
distributions for all 
model parameters and latent variables. \cite{Carlin00} detail the conditions
necessary for the convergence of such a hybrid algorithm. \\

Detailed derivations of the full conditional distributions for the DPPCA model
parameters and latent variables are given in the Supplementary Material. For
the Metropolis-Hastings steps to update the log volatilities, proposal
distributions which are closely related to the shape and orientation of the
target full conditional distributions provide an improved rate of convergence.
To achieve this, second order Taylor expansions of the full conditional
distributions for $\bfeta$ and $\it{\Lambda}$ are employed to guide the
choice of an effective proposal distribution and its parameter values
\citep{kim&shephard98}. A summary of one sweep of the Metropolis-within-Gibbs
sampler for the DPPCA model is given in the Supplementary Material. 
\subsection{Model Identification}
\label{sec:modelidentification}

As with factor analytic models, the DPPCA model suffers from
identification issues. Subjecting the loadings matrix and latent
scores to an orthogonal rotation gives rise to the same distribution
for the observed data. Thus it is not possible to identify the model
parameters from the observed data unless restrictions are imposed.\\

Many attempts to deal with non-identifiability of the related factor analytic
models are detailed in the literature. Most commonly, a unique model is defined
by constraining the loadings matrix such that the first $q$ rows are
lower-triangular with positive diagonal elements \citep{geweke1996}. However
imposing this structure also imposes structure on the ordering of the variables
\citep{aguilar00}. Within the context of the motivating metabolomics
application, such a structure cannot be imposed  on the variables as the
ordering of the spectral peaks within a metabolomics spectrum is important.\\

The approach taken here is to estimate a fully unconstrained loadings matrix
using the Metropolis-within-Gibbs sampler detailed in the Supplementary Material.
Procrustean techniques \citep{Borg05} are then employed to post-process the
sampled loadings matrices to match them to the maximum likelihood estimate (MLE)
of the loadings matrix resulting from fitting a PPCA model to data from the
relevant time point. The MLE is used only as a template, to identify the model.
The transformation required to match the loadings matrices is also applied to
the latent scores. In practice, this has proved to be a fast and satisfactory
approach to dealing with model non-identifiability.

\section{Results }
\label{sec:results}

As detailed in Section \ref{sec:data}, three specific issues associated with the longitudinal metabolomics study need to be addressed: (i) data visualisation, (ii) assessing the effect of time within
each treatment group and (iii) identifying the specific metabolites which change over time within each
treatment group. The DPPCA model, in combination with linear mixed
models, is fitted to the longitudinal metabolomics data set to address these
issues. For reasons of visual clarity, only models with $q = 2$ were considered. For each set of results detailed below, the prior distributions employed for the
DPPCA model parameters were specifically:
\begin{eqnarray*}
\mathbf{w}_{km} & \sim & \mbox{MVN}_{q}(\mathbf{0}, \emph{I})
\hspace{0.25cm} \mbox{ for } k = 1, \ldots, p \mbox{ and } m = 1, \ldots, M.\\
\nu & \sim &N(0, 10)\\
v^{2} & \sim & IG(6/2, 0.5/2)\\
\phi & \sim & N_{[-1, 1]}(0.75, 0.1)
\end{eqnarray*}
The priors on the univariate entries of the set of parameters $\theta_{2} =
(\bmu, \it{\Phi}, \it{V})$ were the same as those for
$\theta_{1} = (\nu, \phi, v^{2})$. The Metropolis-within-Gibbs sampler was run
for 500,000 iterations, thinned every $500^{th}$ iteration. The first 5,000
iterations were discarded as burn-in. The
MCMC algorithm was initialized using estimates of the loading matrices from
fitting a PPCA model to data from each time point independently; stochastic
volatility model parameters were set equal to their prior means. Trace plots and
autocorrelation function (ACF) plots for the MCMC samples of the parameters were
used to assess convergence of the algorithm.

\subsection{Data Visualisation: Exploring Metabolomic Trajectories}

In longitudinal metabolomics studies, trajectories through the latent principal
subspace can be used to gain visual insight to the response of animals during the
study period. Examining the location, magnitude and direction of these metabolomic
trajectories provides visual insight to the metabolomic changes over time.\\

Here metabolomic trajectories were estimated using the latent scores
of animals resulting from collectively modelling data from both treatment groups using a DPPCA model.  Such a
model takes into account the covariation between the metabolites and
any correlation across time; this facilitates visualisation of animals in a reduced dimensional space, while appropriately
modelling the time course nature of the data. Trace plots for the estimated
latent scores and loadings are given in the Supplementary Material.\\

\begin{figure}[h!]
\begin{center}
\begin{tabular}{cc}
\hspace{-1cm}\subfigure[]{
\includegraphics[width=7cm, height=7cm]{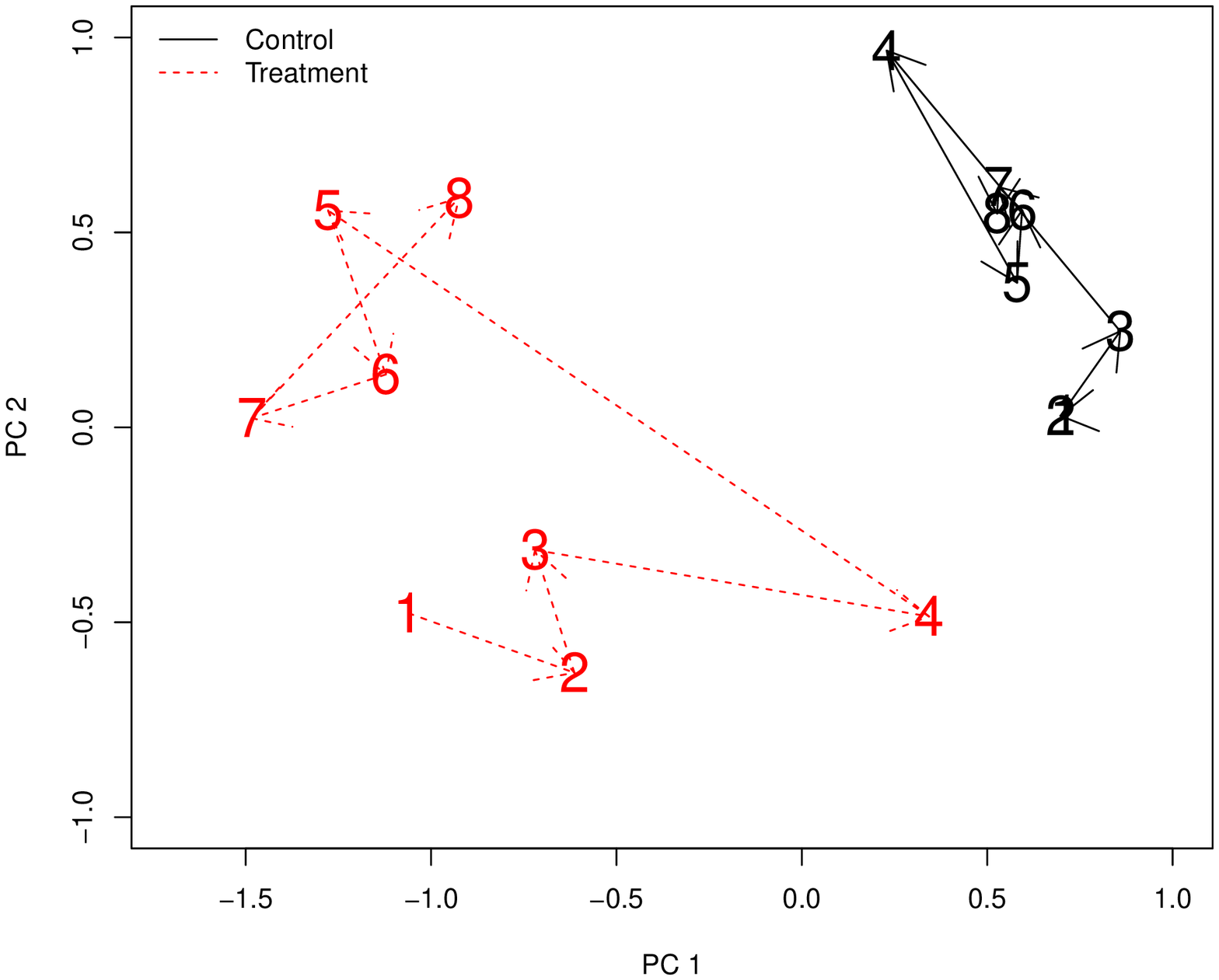}}&
\subfigure[]{
\includegraphics[width=7cm, height=7cm]{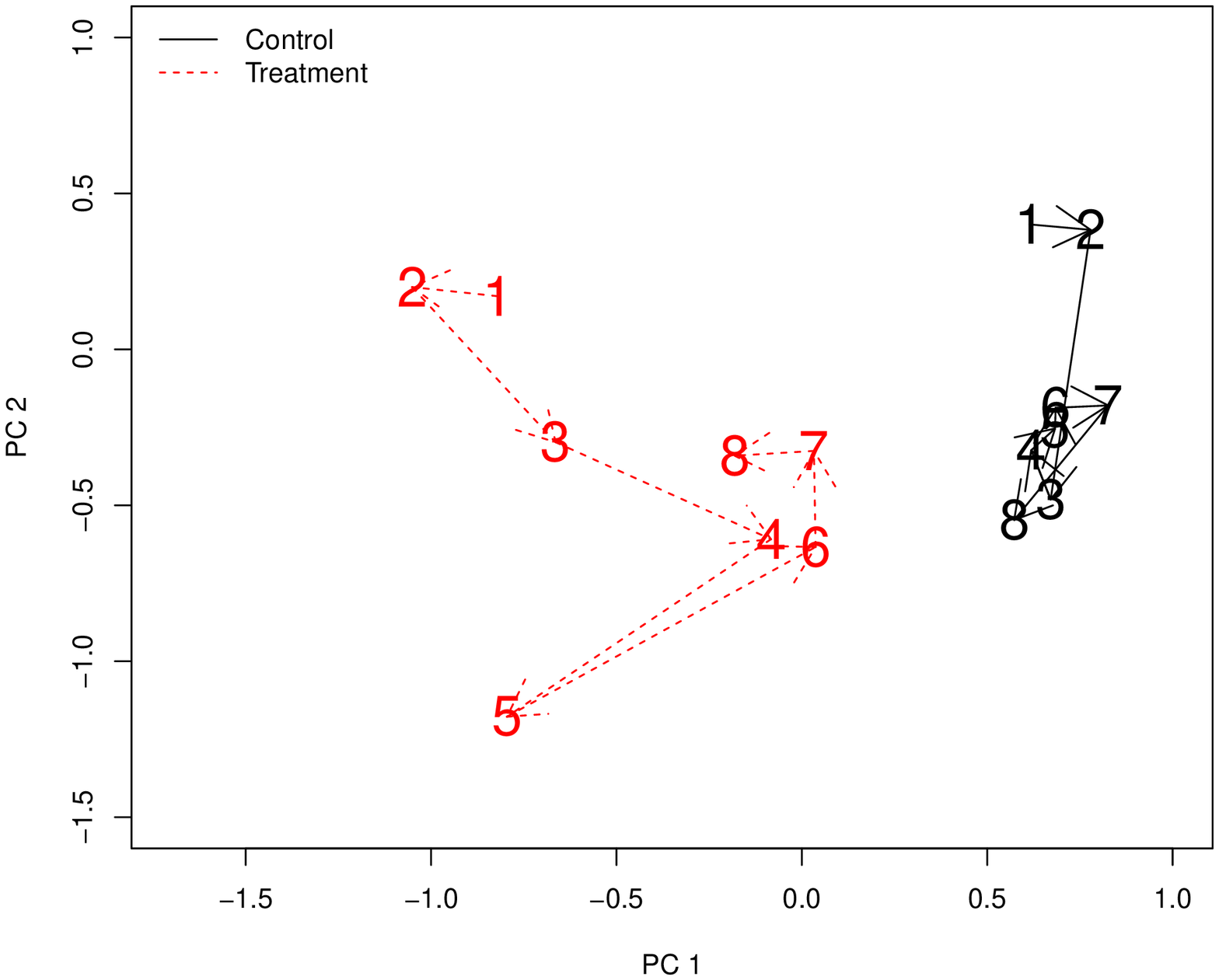}}
\end{tabular}
\end{center}
\caption{\label{fig:trajectories}Individual trajectories for four randomly
sampled animals, in the principal subspace from the first time point. (a) An
animal from the control group (black solid lines) and an animal from the treated
group  (red dashed lines) and (b) an animal from the control group (black solid
lines) and an animal from the treated group (red dashed lines). The digits
represent the time points of the study and arrows illustrate movement through
time.}
\end{figure}

The metabolomic trajectories of four randomly sampled animals are
illustrated in Figure \ref{fig:trajectories}. Under the DPPCA model, each time point $m$ has a different principal subspace, defined by the columns of the relevant loadings matrix $\emph{W}_{m}$. Hence the latent scores of animals at different time points lie in different subspaces. To visualise the metabolomic trajectories the latent scores must therefore be unified. This is achieved by again drawing on Procrustean ideas, where the loadings matrix from the first time point is used as the reference matrix. The loadings matrix from each subsequent time point $m$ is rotated to
best match the loadings matrix from the first time point; the same rotation is then applied to
the associated set of scores from time point $m$. This facilitates illustration of the movement of the
latent scores over time within the same principal subspace.
Figure \ref{fig:trajectories} therefore provides visual insight to the animals' metabolomic trajectories in the principal subspace from the
first time point.\\

Figure \ref{fig:trajectories} suggests the presence of a treatment effect through
the visible separation of the locations of the treated and control animals in
the principal subspace from the first time point.  The difference in the
biochemical composition of the urine due to treatment is highlighted by the
different `metabolic starting positions' of the trajectories for the randomly
selected animals from the control group and those from the treatment groups.
This is due to the fact that the urine samples analysed at time point 1 actually
resulted from day 3 of the study, at which stage the treatment is
apparently having an effect.\\

The trajectories also demonstrate that the magnitude of the metabolic changes in
the biochemical composition of the urine samples is much greater in the
treatment group than in the control group, over time. This is evidenced by the
larger movements between time points by the treated animals. This shows that the
variability in the urinary composition of the treated animals over time is
greater than that in the control group. Thus, the metabolomic trajectories provide a visual insight to the metabolomic changes occurring over time. \\

\subsection{Exploring the Effect of Time}

The second aim of the longitudinal study was to ascertain if there is a time effect within each treatment group.
In an effort to quantify the effect of time, the DPPCA model was fitted
separately to each treatment group. If a time effect is established, the task
will then be to identify metabolites whose concentration level is significantly
changing over time.

\subsubsection{Exploring the Effect of Time in the Treatment Group}
\label{sec:treatmentgroup}

\begin{figure}[h]
\centering
\begin{tabular}{ccc}
\subfigure[]{\includegraphics[width=0.3\textwidth,
height=50mm]{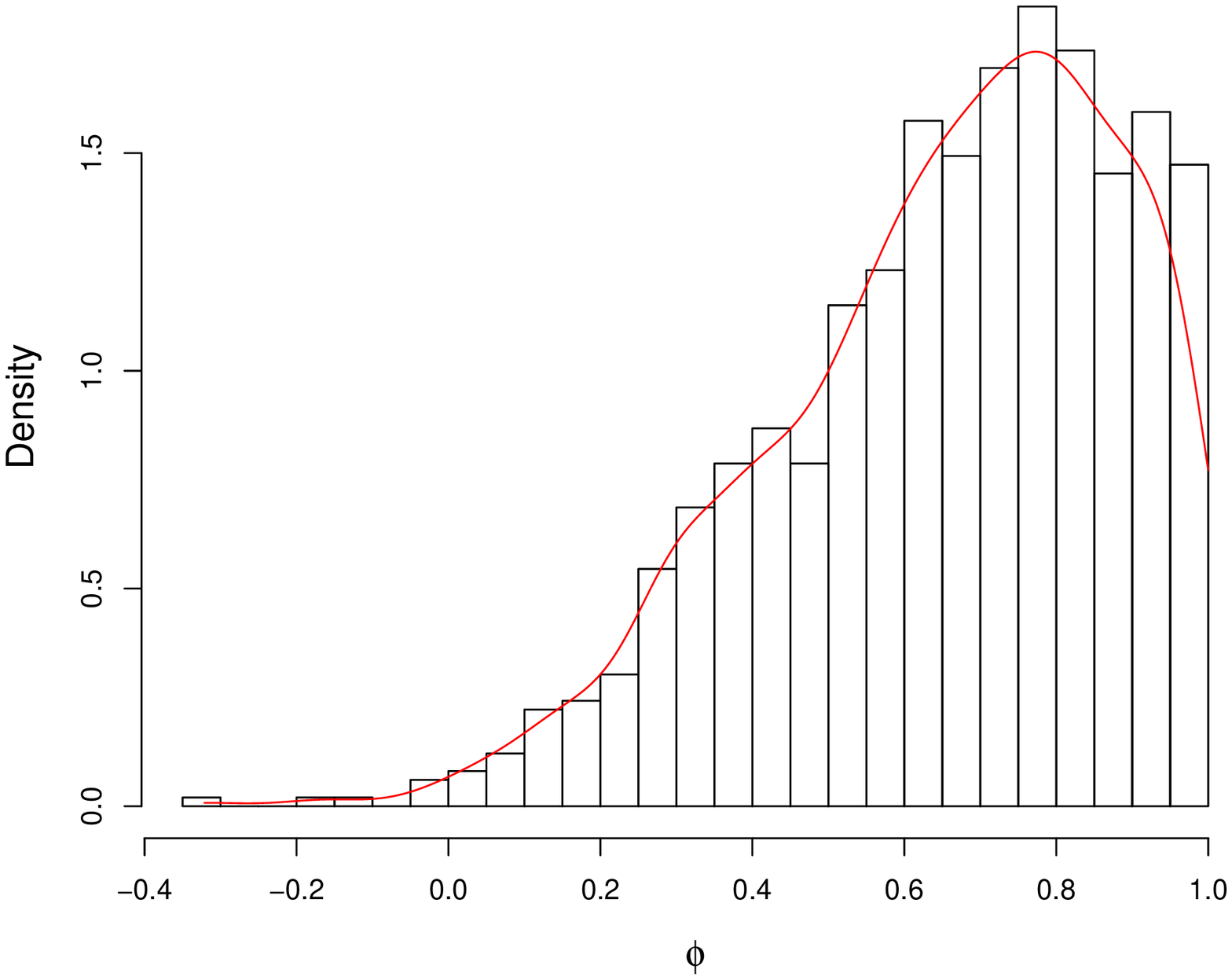}
   \label{fig:phiden}}&
\subfigure[]{\includegraphics[width=0.3\textwidth, height=50mm]{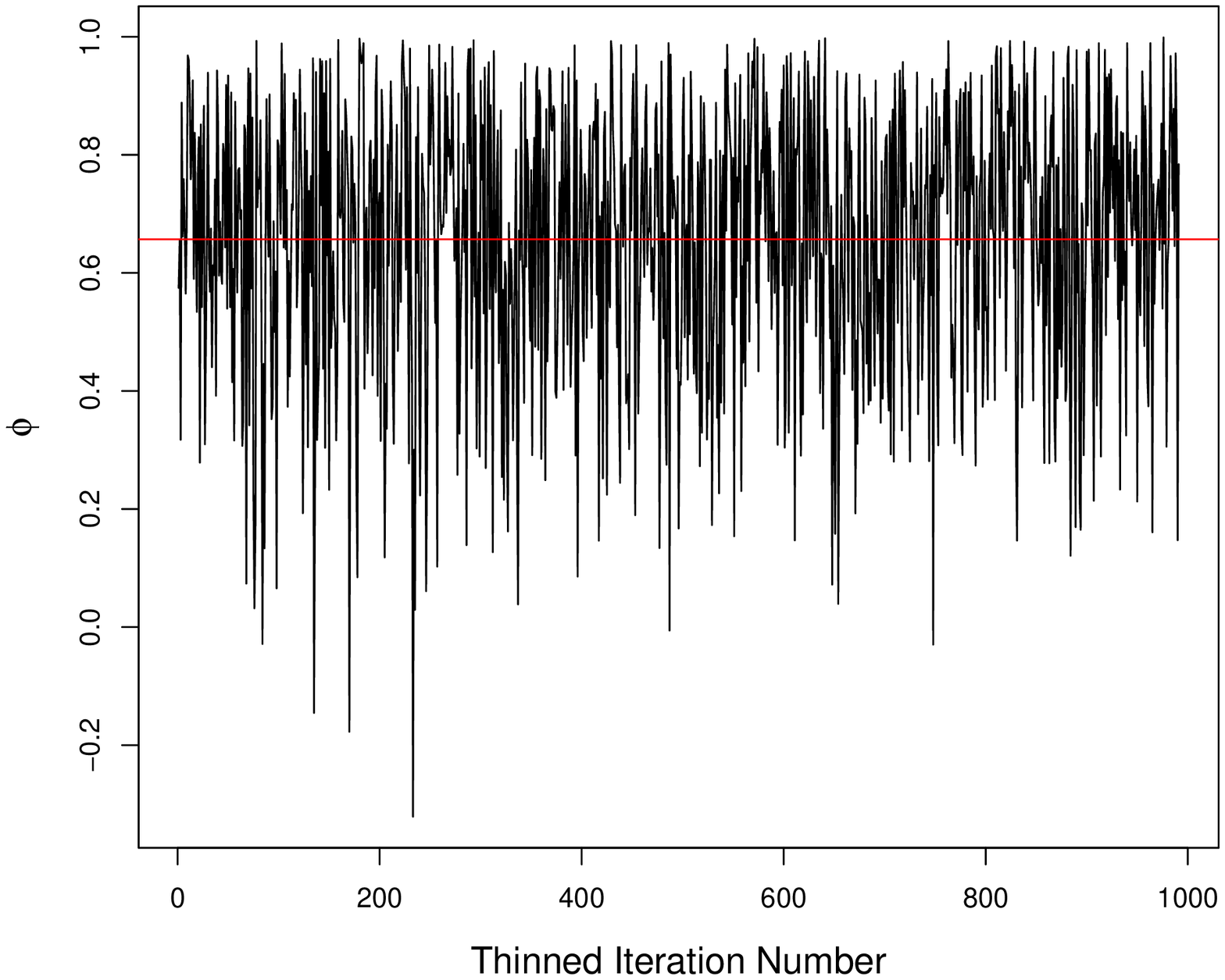}
\label{fig:phitrace}}&
\subfigure[]{\includegraphics[width=0.3\textwidth,
height=50mm]{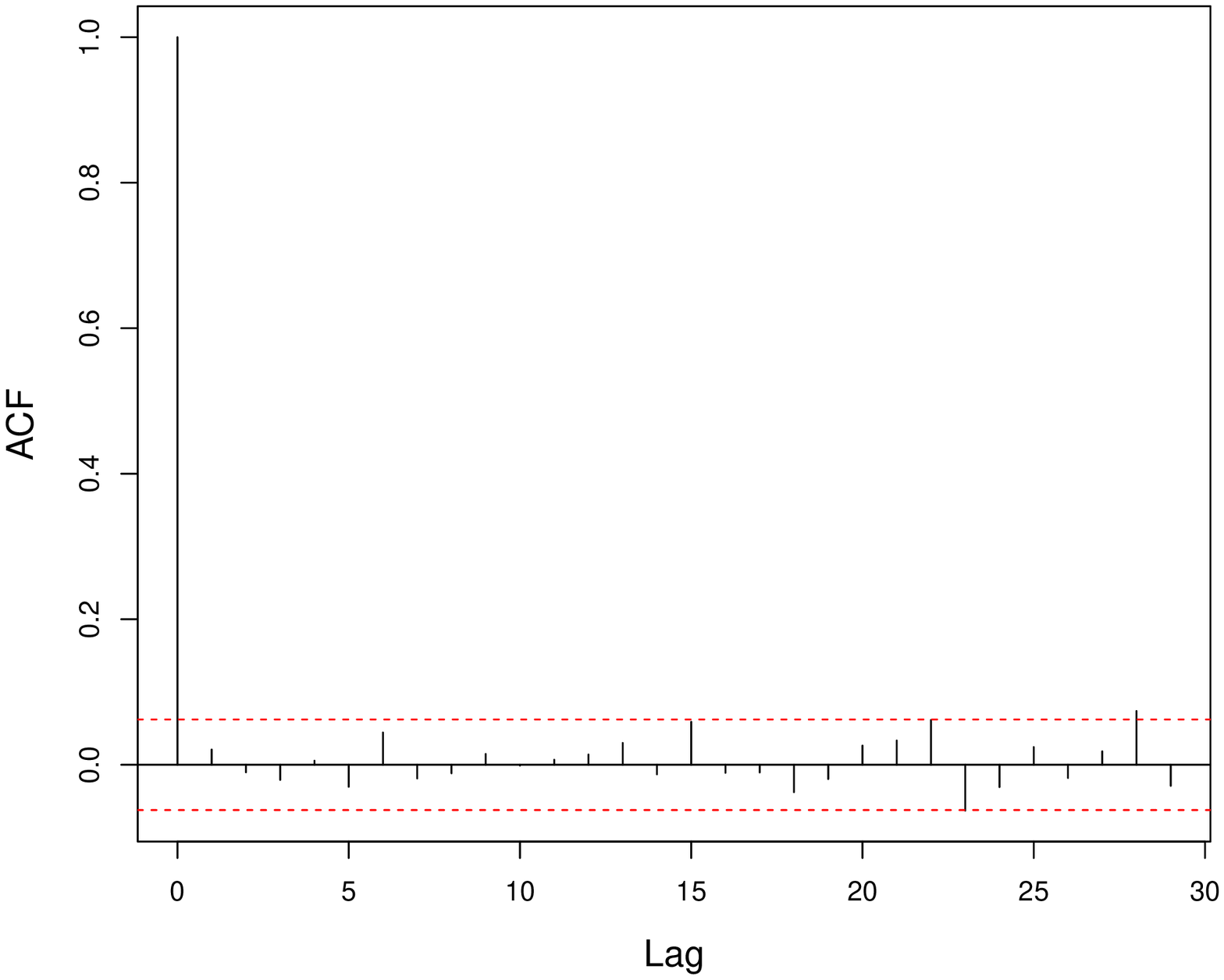}  \label{fig:phiacf}}
\end{tabular}
\caption{\label{fig:phi}The persistence parameter, $\phi$, of the SV model on
the error variances in the treatment group: (a) plot of the posterior density,
(b) trace plot and (c) ACF plot. The horizontal line in (b) illustrates the
posterior mean of $\phi$.}
\end{figure}

The DPPCA model was fitted to the metabolomic spectra from the animals in the
treatment group. The persistence parameters in the SV models are the parameters
of interest as they quantify the strength of the relationship between the time
points. Figure \ref{fig:phiden} illustrates the posterior distribution of the
persistence parameter ($\phi$) of the SV model on the errors. The
relevant trace and ACF plots are given in Figure \ref{fig:phitrace} and Figure
\ref{fig:phiacf} respectively. The posterior mean of $\phi$ was large and
positive ($\hat{\phi} = 0.69$) and significant (95$\%$ quantile based credible interval (CI)
(0.15, 0.97)). The persistence parameters of the SV model on the latent variables for
PC 1 and PC 2 were also estimated to be large and significant at $\hat{\phi}_{1}
= 0.64$ (0.07, 0.97) and $\hat{\phi}_{2} = 0.66$ (0.08, 0.97), respectively. The
posterior means suggest that a positive time dependency exists among the spectra
from the treatment group.\\

Given that a time effect has been established, the third aim of the study was to identify the specific metabolites which change over time within the 
treatment group. This is achieved by first using the DPPCA model to
expose those metabolites which influence the data structure at each time point.
Under the DPPCA model, this translates to identifying a subset of metabolites whose posterior mean
loadings are largest (in terms of magnitude) at each time point. Standard
linear mixed models are then fitted to these `influential metabolites' to
identify those which change over time. This approach yields a panel of
metabolites which evolve over time, while appropriately accounting for the
covariation in the high-dimensional data, and the time related dependencies.\\

\begin{figure}[!hp]
\centering
\begin{tabular}{ccc}
\includegraphics[width=0.35\textwidth,height=0.19\textheight]{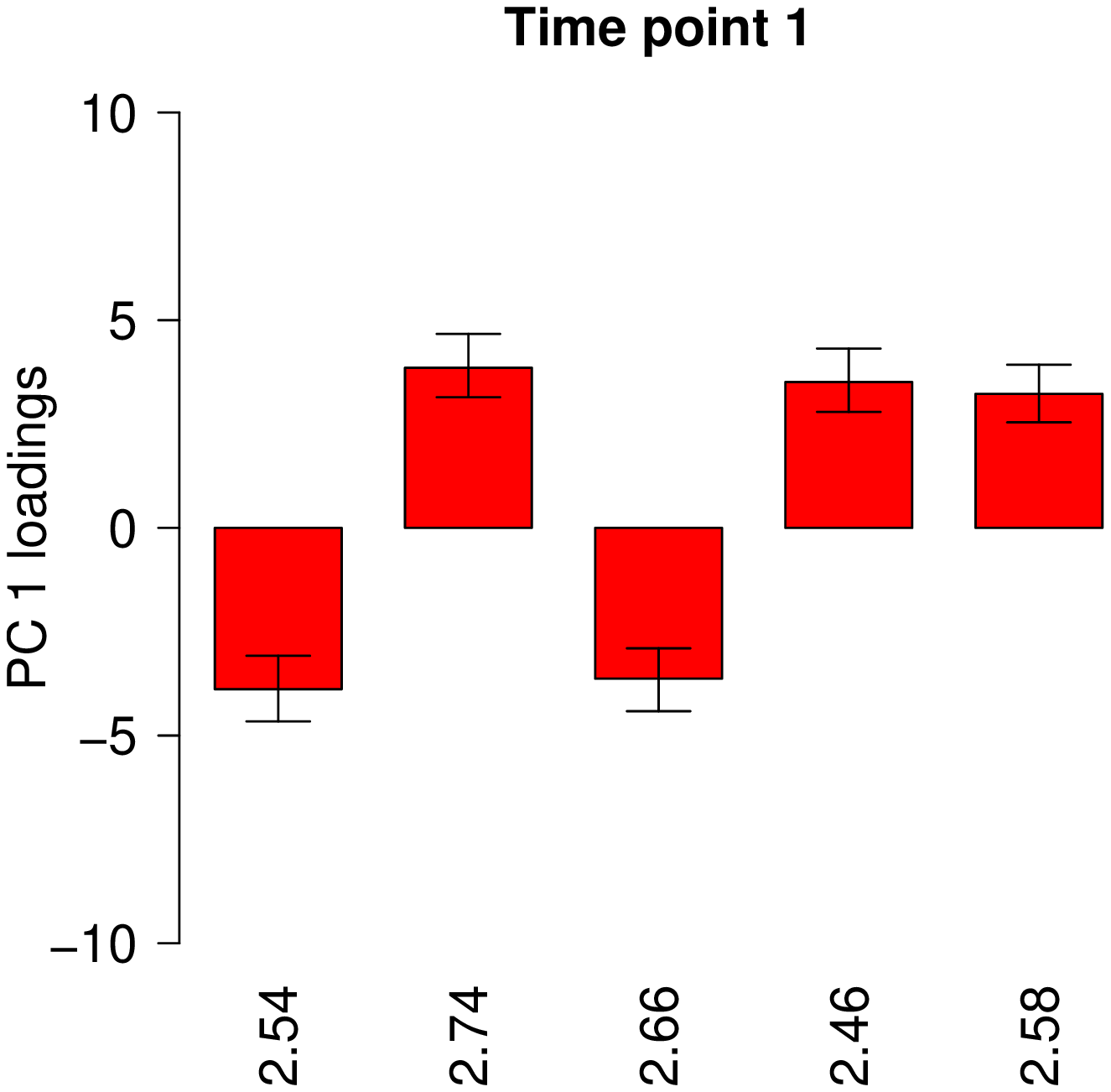}
& \hspace{0.1cm} &
\includegraphics[width=0.35\textwidth,height=0.19\textheight]{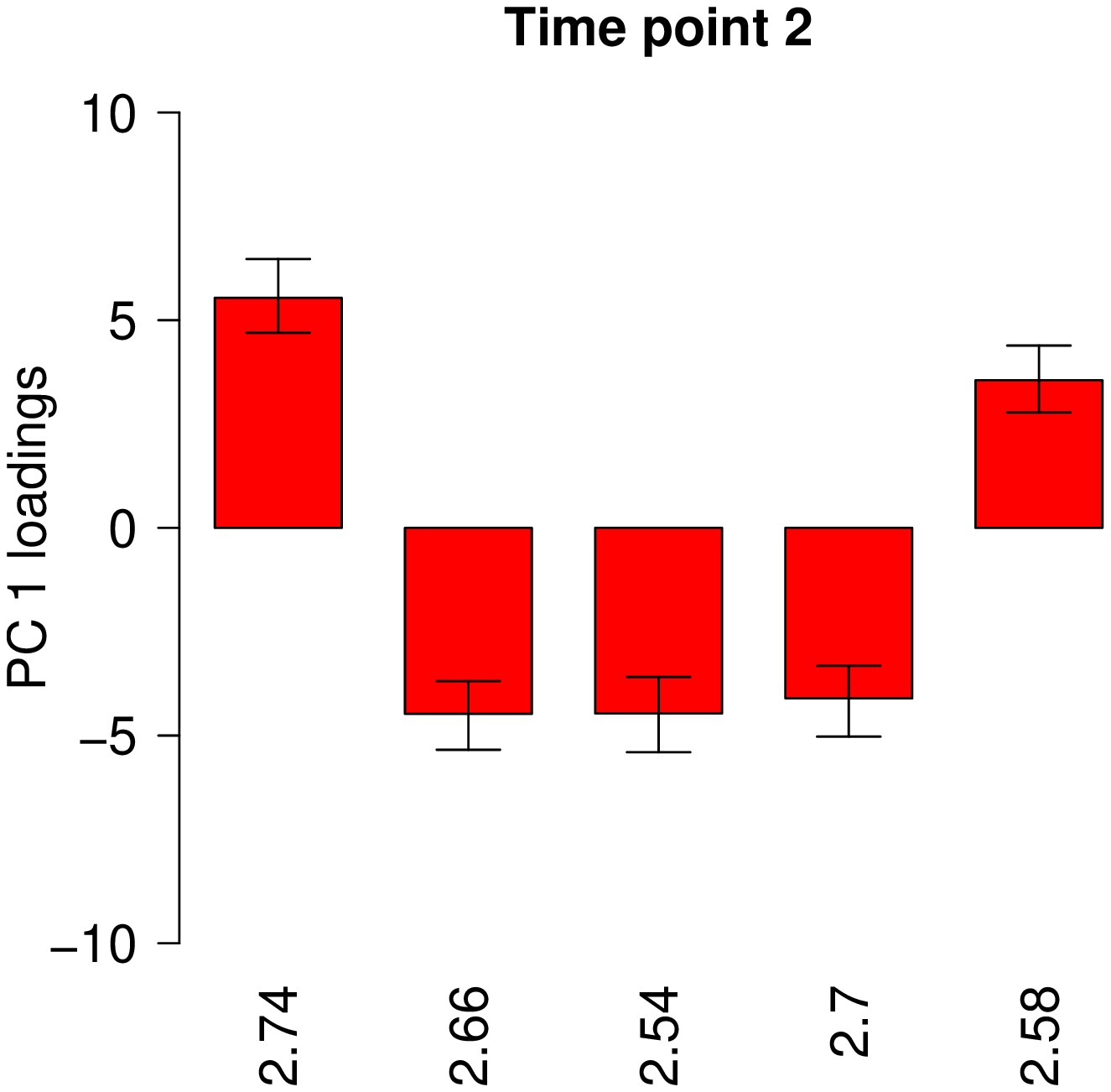}
\\
 & & \\
\includegraphics[width=0.35\textwidth,height=0.19\textheight]{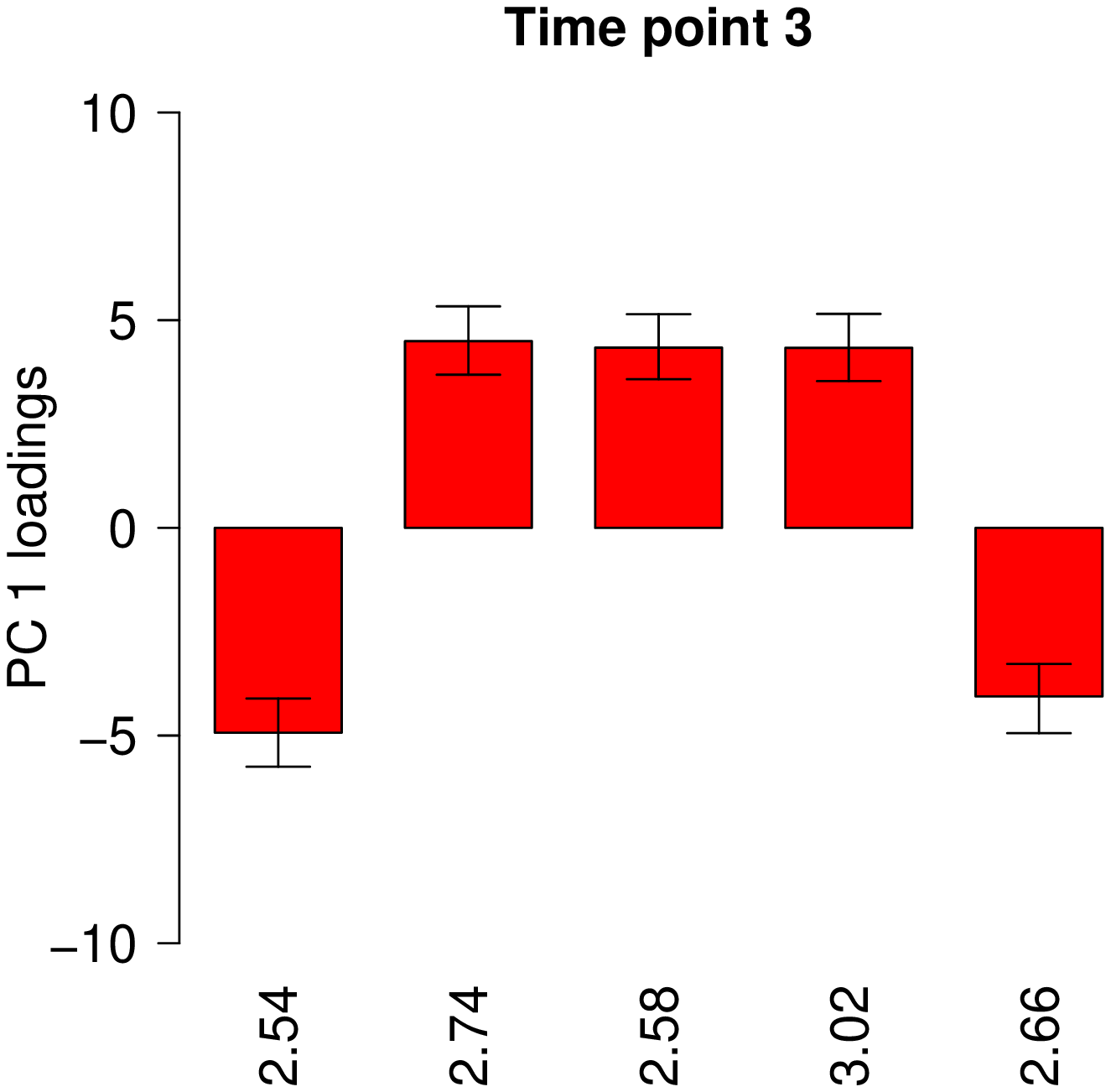}
&\hspace{0.1cm} &
\includegraphics[width=0.35\textwidth,height=0.19\textheight]{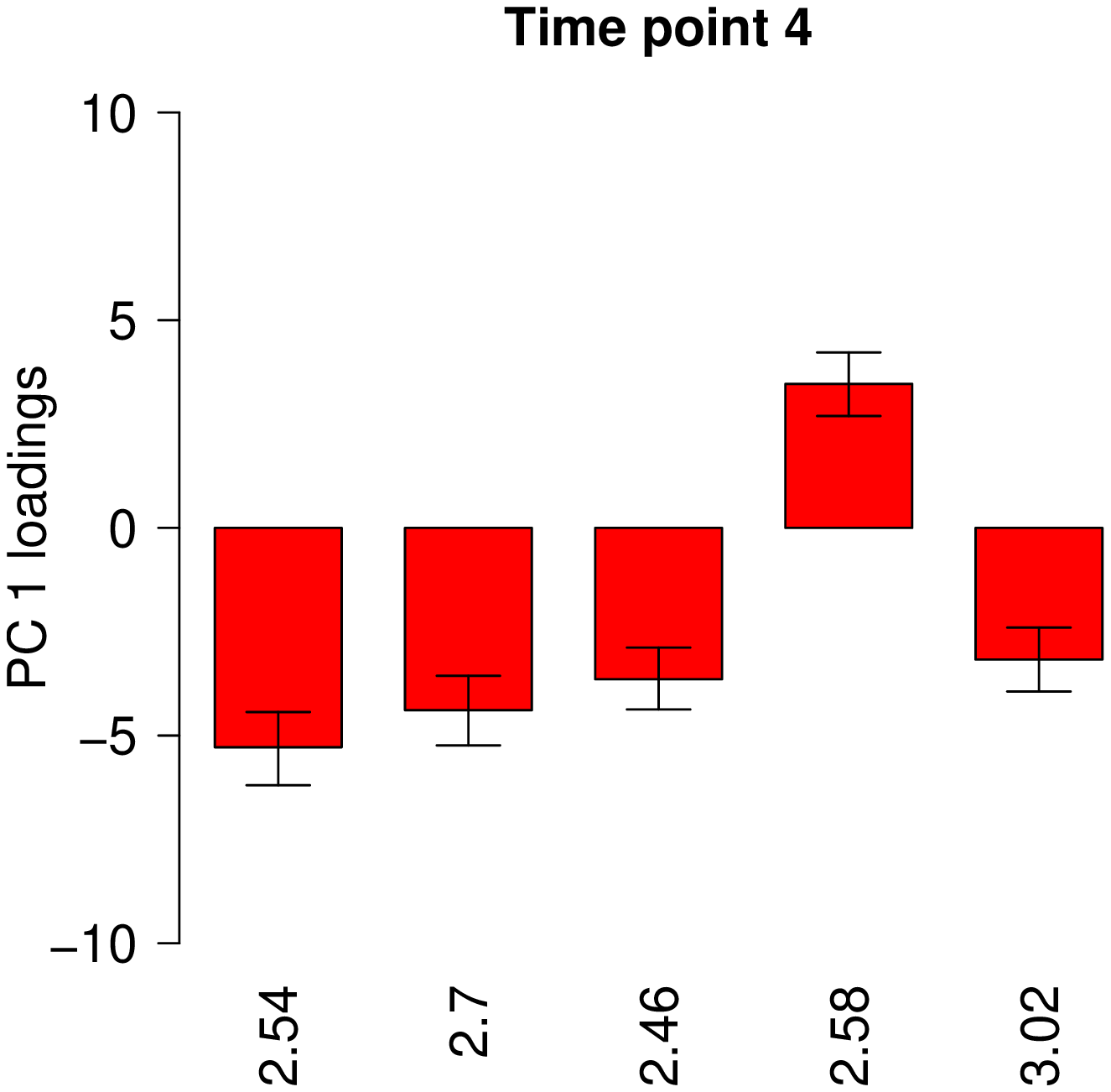}
\\
 & & \\
\includegraphics[width=0.35\textwidth,height=0.19\textheight]{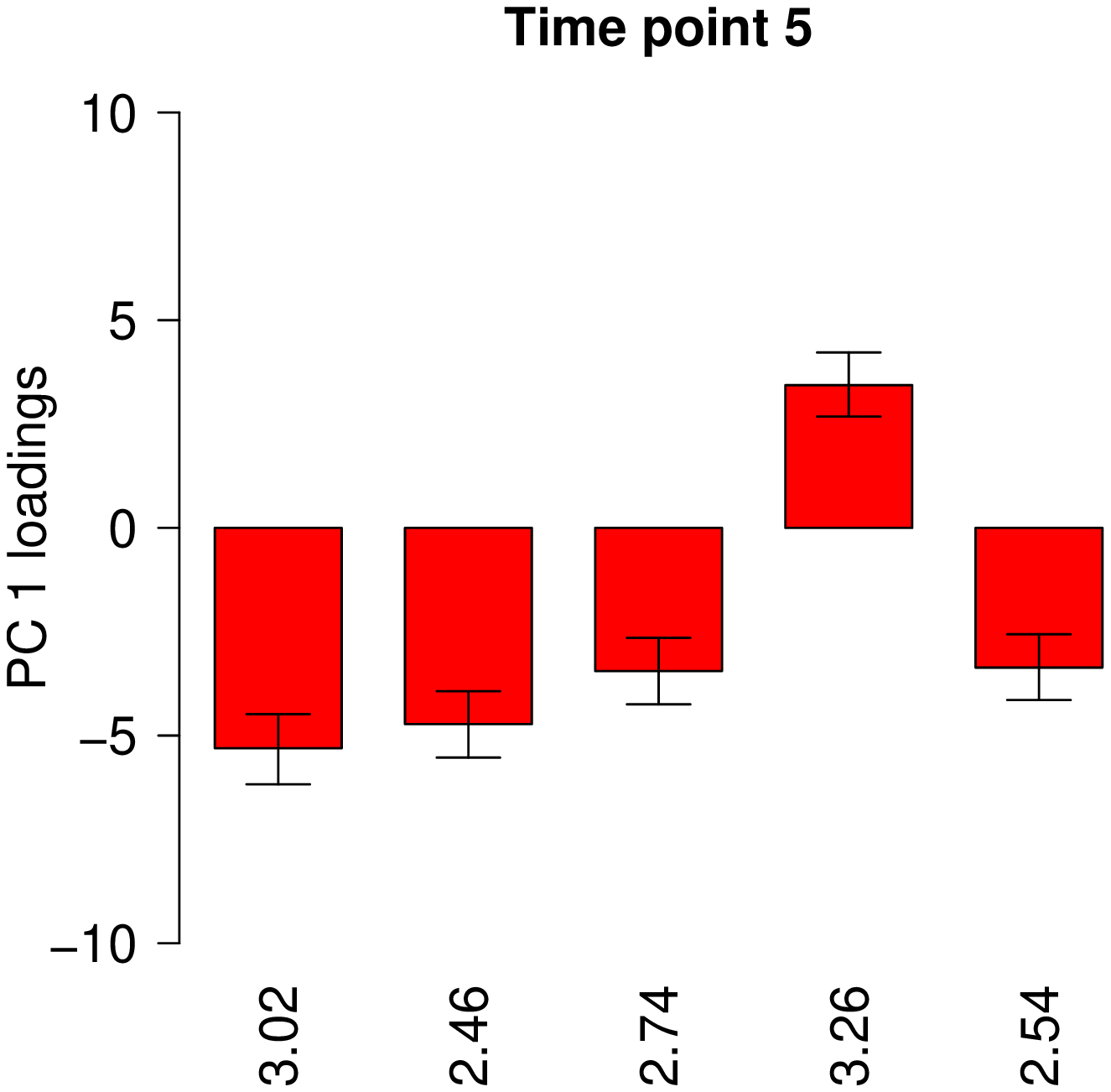}
&\hspace{0.1cm} &
\includegraphics[width=0.35\textwidth,height=0.19\textheight]{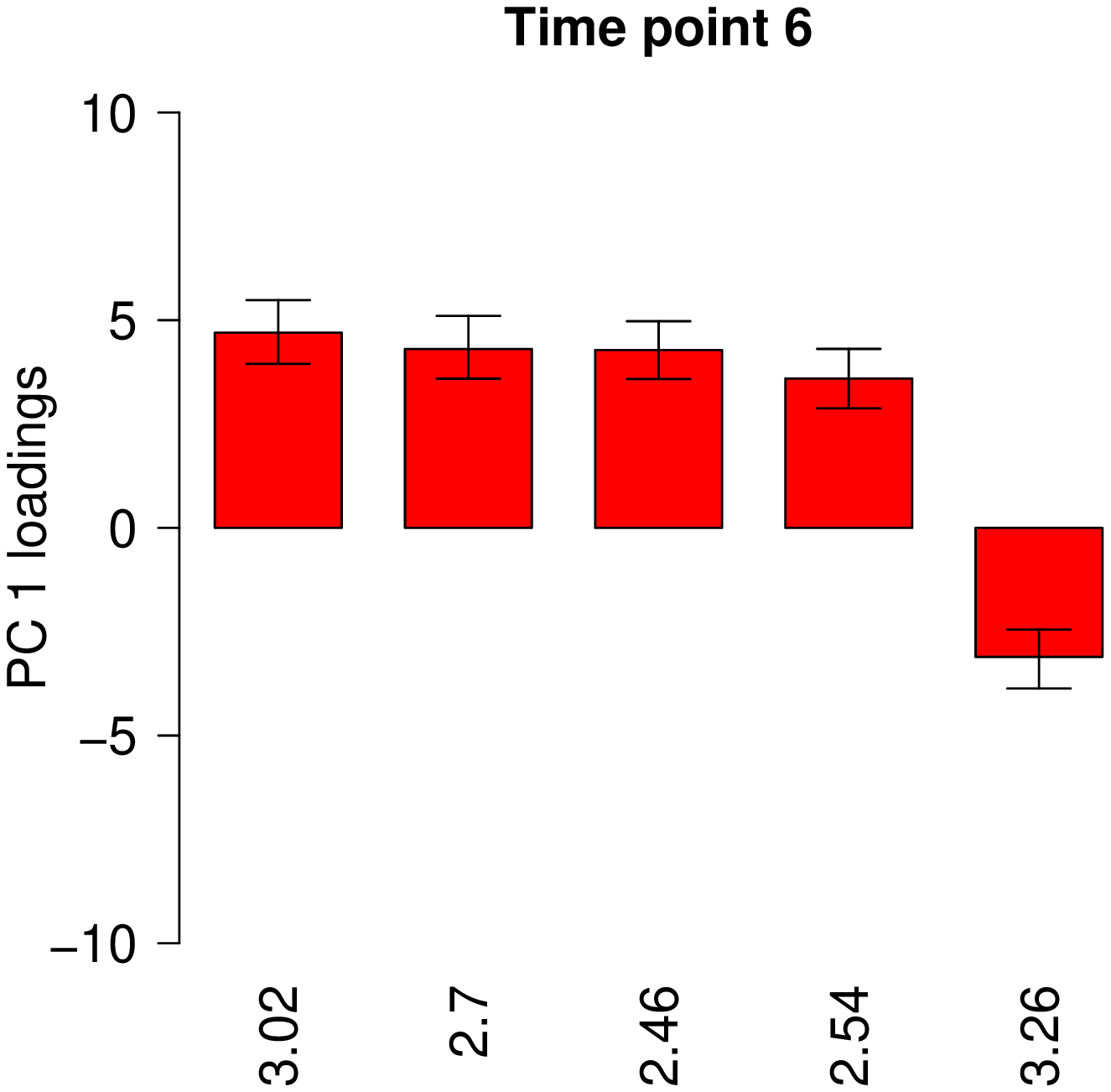}
\\
 & & \\
\includegraphics[width=0.35\textwidth,height=0.19\textheight]{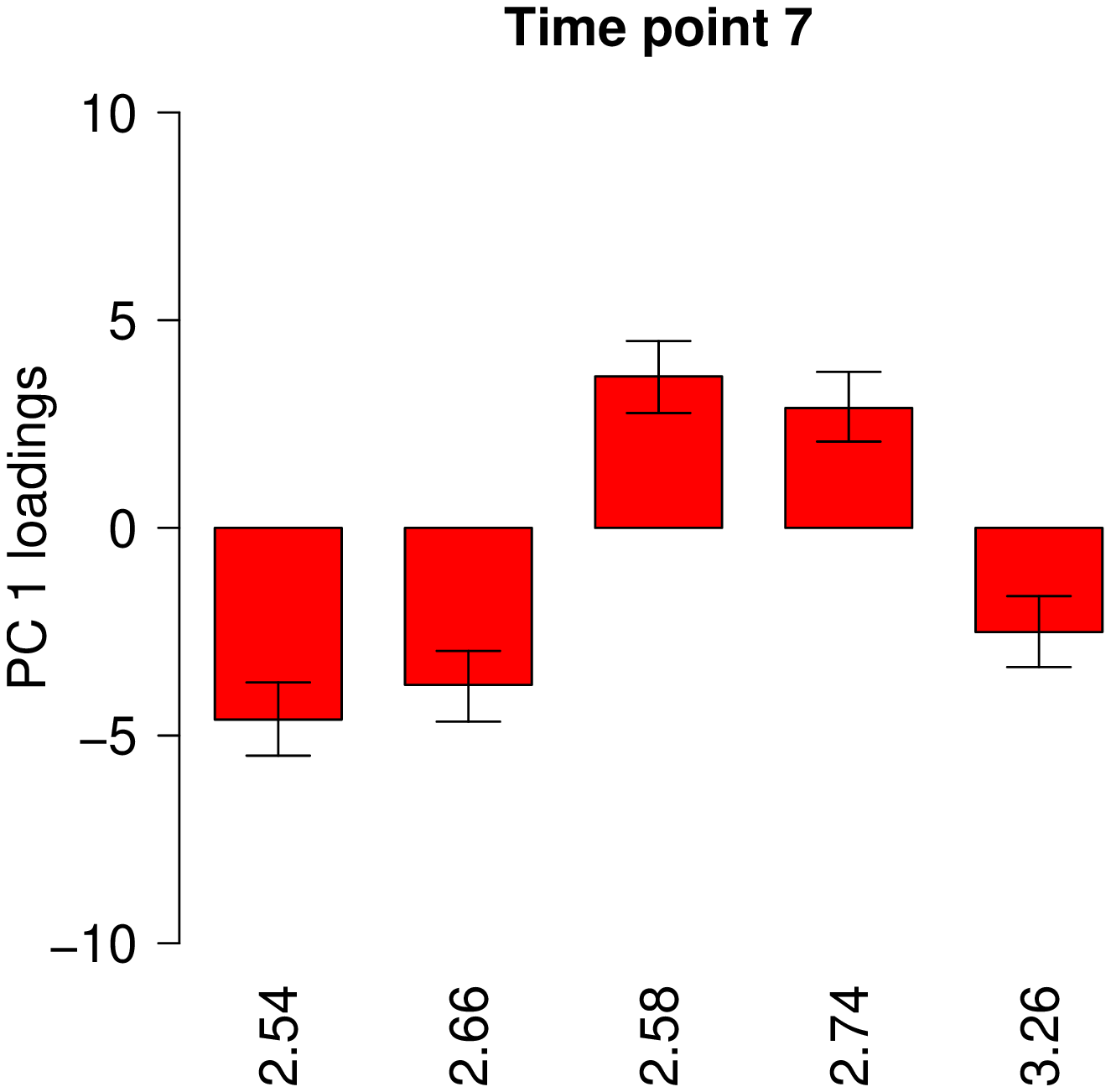}
& \hspace{0.1cm} &
\includegraphics[width=0.35\textwidth,height=0.19\textheight]{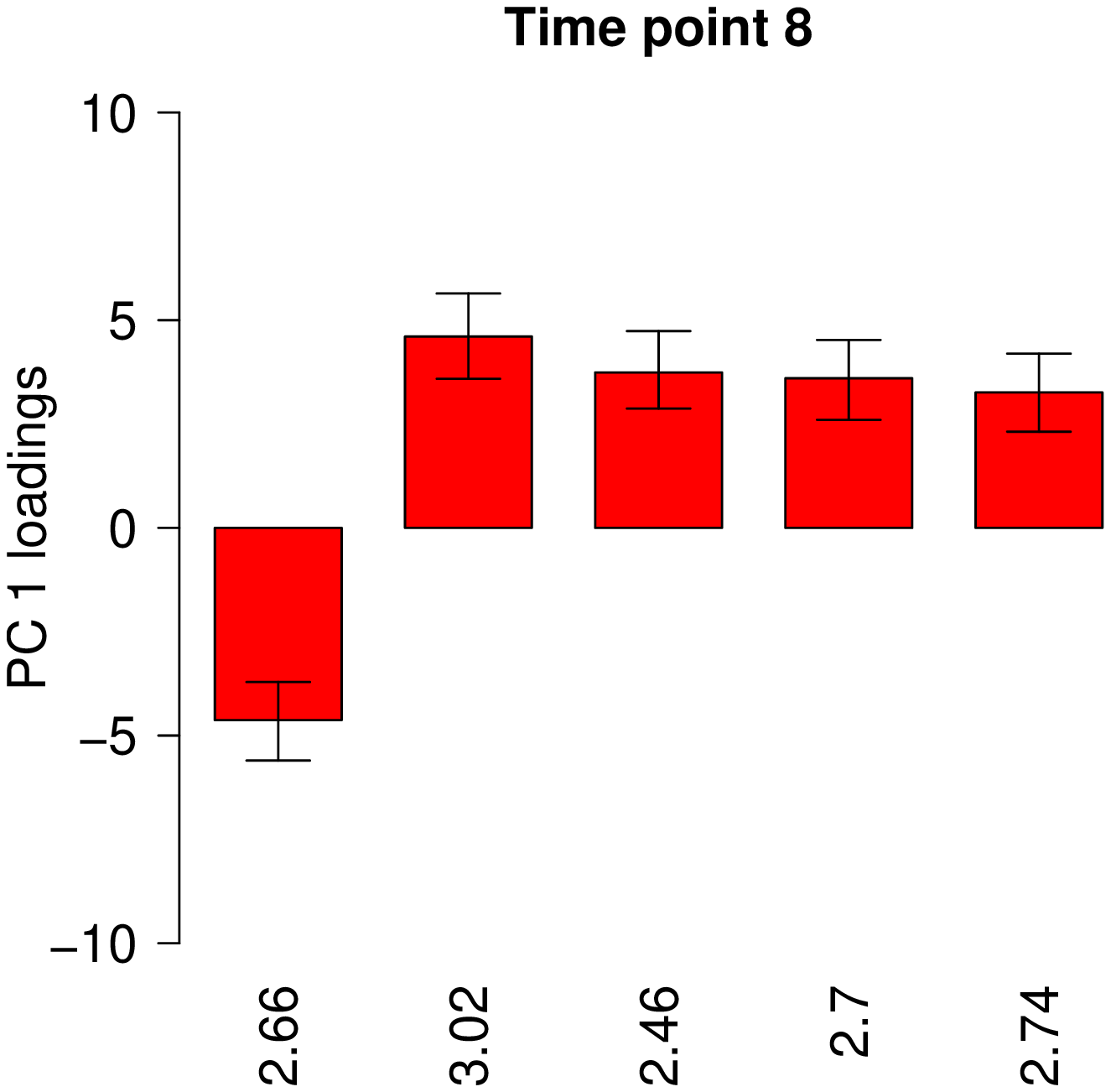}
\\
\end{tabular}
\caption{\label{fig:barplots}Barplots of the posterior mean loadings for the top
five influential spectral bins, which correspond to metabolites, in the
treatment group. The error bars are the corresponding 95\% quantile based credible intervals.}
\end{figure}

After fitting the DPPCA model to the spectra from animals in the treatment group, several spectral regions (corresponding to metabolites) were identified as influencing the underlying structure of the data.  At each time point, the absolute values of the posterior mean loadings on PC1 were ranked in descending order. The top five influential spectral bins at each time point were determined and are shown in Figure \ref{fig:barplots}. None of the 95\% CIs associated with these spectral bins included zero. The set of the top five spectral bins across all $M = 8$ time points consists of only eight unique spectral bins (2.46ppm, 2.54ppm, 2.58ppm, 2.66ppm, 2.7ppm, 2.74ppm, 3.02ppm and 3.26ppm). \\

\begin{figure}[h]
\centering
\includegraphics[width=0.65\textwidth,height=0.4\textheight]{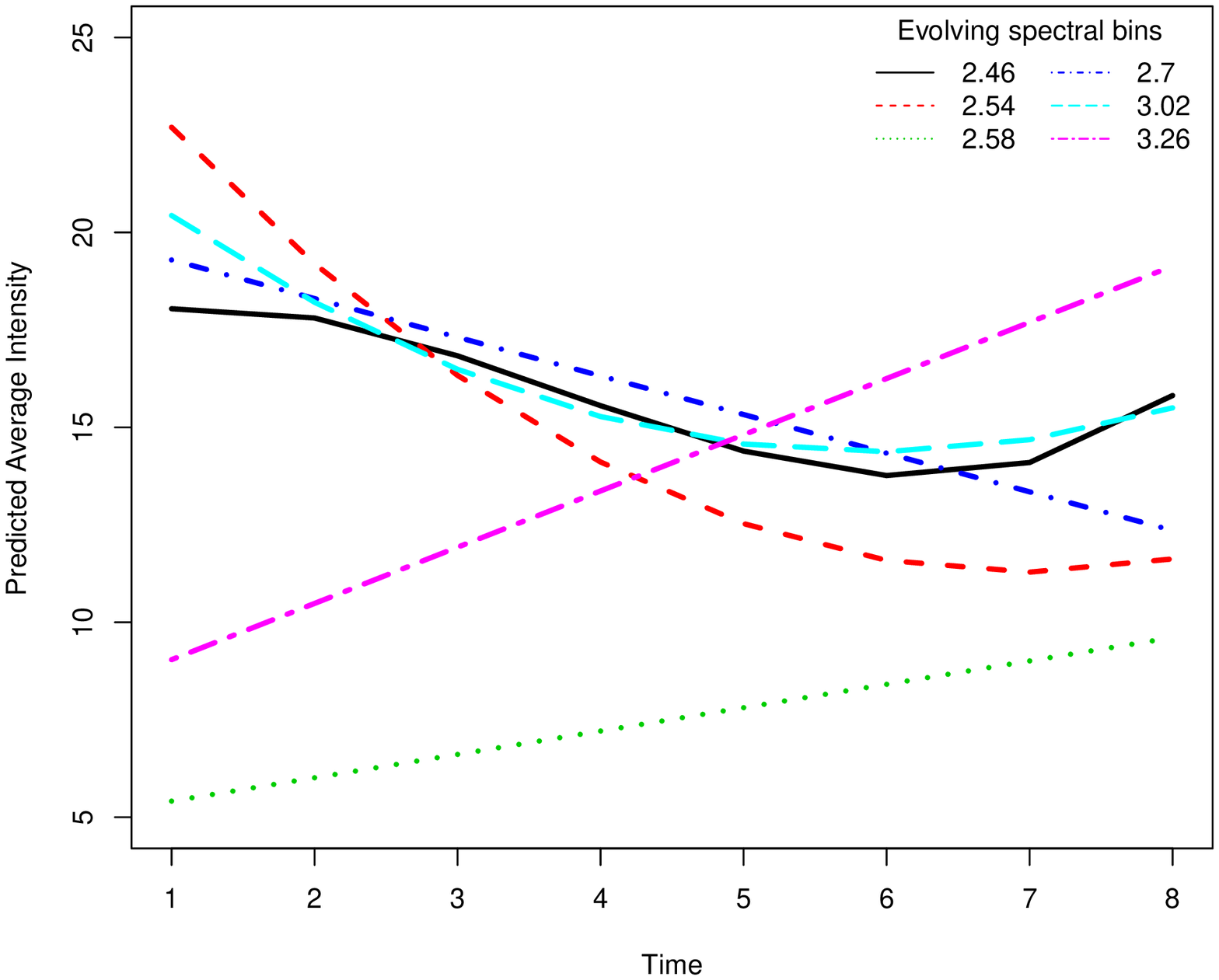}
\caption{\label{fig:lmmtreat}The LMM predicted average intensities of the six
influential spectral bins which evolve over time in the treatment group.}
\end{figure}

Bayesian linear mixed models were fitted to the data associated with the eight
unique influential spectral bins to determine which, if any, have concentrations
which evolve over time. A random intercept model with cubic time effect was the
most complex model considered; no interaction terms were considered. A backwards
selection type approach was taken to model selection for each spectral bin
considered. Of the eight spectral bins considered, six were deemed to have
significantly fluctuating concentration levels over time. Figure \ref{fig:lmmtreat}
illustrates the predicted average intensity levels for each of the six spectral
bins.\\

The metabolites identified to be evolving over time include the metabolite
2-oxoglutarate, represented by the spectral bins 2.46ppm and 3.02ppm. The
concentration level of 2-oxoglutarate decreases initially during the study and
increases at later time points, as illustrated by the similar behaviour of the
predicted intensities of 2.46ppm and 3.02ppm in Figure \ref{fig:lmmtreat}. The
model also predicts a linear decreasing metabolic time profile for spectral bin
2.7ppm. Spectral bin 2.54ppm has a positive quadratic time effect in the treated
animals i.e. the concentration level decreases and then increases over time.
Spectral bins 2.58ppm and 3.26ppm have a positive linear time trend. Individual
animal and predicted profiles for three of the six evolving spectral bins are
given in the Supplementary Material.

\subsubsection{Exploring the Effect of Time in the Control Group}

To establish the presence or absence of a time effect in the control group of
animals, and to subsequently highlight those metabolites which evolve over time,
the same approach as that taken in Section \ref{sec:treatmentgroup} was
followed. That is, the DPPCA model was fitted to the spectra of animals in the
control group only; Table \ref{tab:control} details the posterior means of the
persistence parameters of the SV model on the errors and on the latent
variables, with their corresponding 95\% CIs. Table \ref{tab:control}
shows that the persistence parameters of the SV models are large and
significant, suggesting that there is a relationship across time. \\

\begin{table}
\caption{\label{tab:control}Posterior means of the persistence parameters and
the corresponding 95\% CIs for the control group.}
\centering
\fbox{%
\begin{tabular}{l|cl}\hline\hline
SV model &Estimate & (95$\%$ CI) \\\hline
Errors ($\phi$)  & 0.66 & (0.09,0.98) \\
PC 1 ($\phi_1$) & 0.65 &(0.10,0.98) \\
PC 2 ($\phi_2$) & 0.66 &(0.07,0.97)\\\hline
\end{tabular}}
\end{table}

\begin{figure}[h]
\centering
\includegraphics[width=0.65\textwidth,height=0.4\textheight]{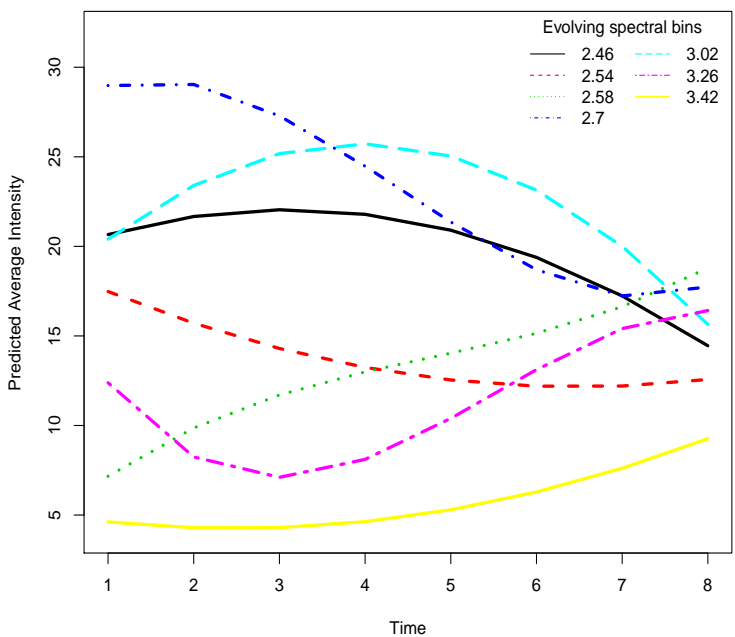}
\caption{\label{fig:lmmcon} The LMM predicted average intensities of the seven
influential spectral bins which evolve over time in the control group.}
\end{figure}

Given that a time effect has been established in the control group, interest
then lies in highlighting those metabolites which evolve over time. The posterior mean PC1 loadings of the DPPCA model were ranked to select the top five
influential spectral bins at each time point; again, none of the associated 95\% CIs included zero. From this list of spectral bins,
those which evolve over time in the control group were identified. Seven unique
influential spectral bins were ranked in the top five over the eight time
points; Bayesian LMM models were fitted to the profiles for each of these and
all seven were identified as evolving over time. Figure \ref{fig:lmmcon}
illustrates the predicted average intensity levels over the eight time points,
under the selected LMM for each of the seven evolving spectral bins.\\

The metabolite 2-oxoglutarate (with corresponding spectral bins 2.46ppm and
3.02ppm) was predicted by the Bayesian LMM to have a negative quadratic time
effect in the control group i.e. its concentration increases and then decreases
over time (see Figure \ref{fig:lmmcon}). Spectral bins 2.54ppm and 3.42ppm have
positive quadratic time effects. The remaining evolving spectral bins (2.58ppm,
2.7ppm and 3.26ppm) have cubic time effects. Individual animal and predicted
profiles for three of the seven evolving spectral bins are given in the Supplementary Material.

\subsection{Comparing evolving metabolites in the two treatment groups}

As the aim of the longitudinal metabolomics study was to determine metabolic
changes that occur over time during PTZ treatment, of interest are the
similarities and differences between the set of evolving metabolites in the
treatment group and the set in the control group. \\

\begin{figure}[h!]
\centering
\begin{tabular}{cc}
\subfigure[]{\includegraphics[width=0.45\textwidth,height=0.225\textheight]{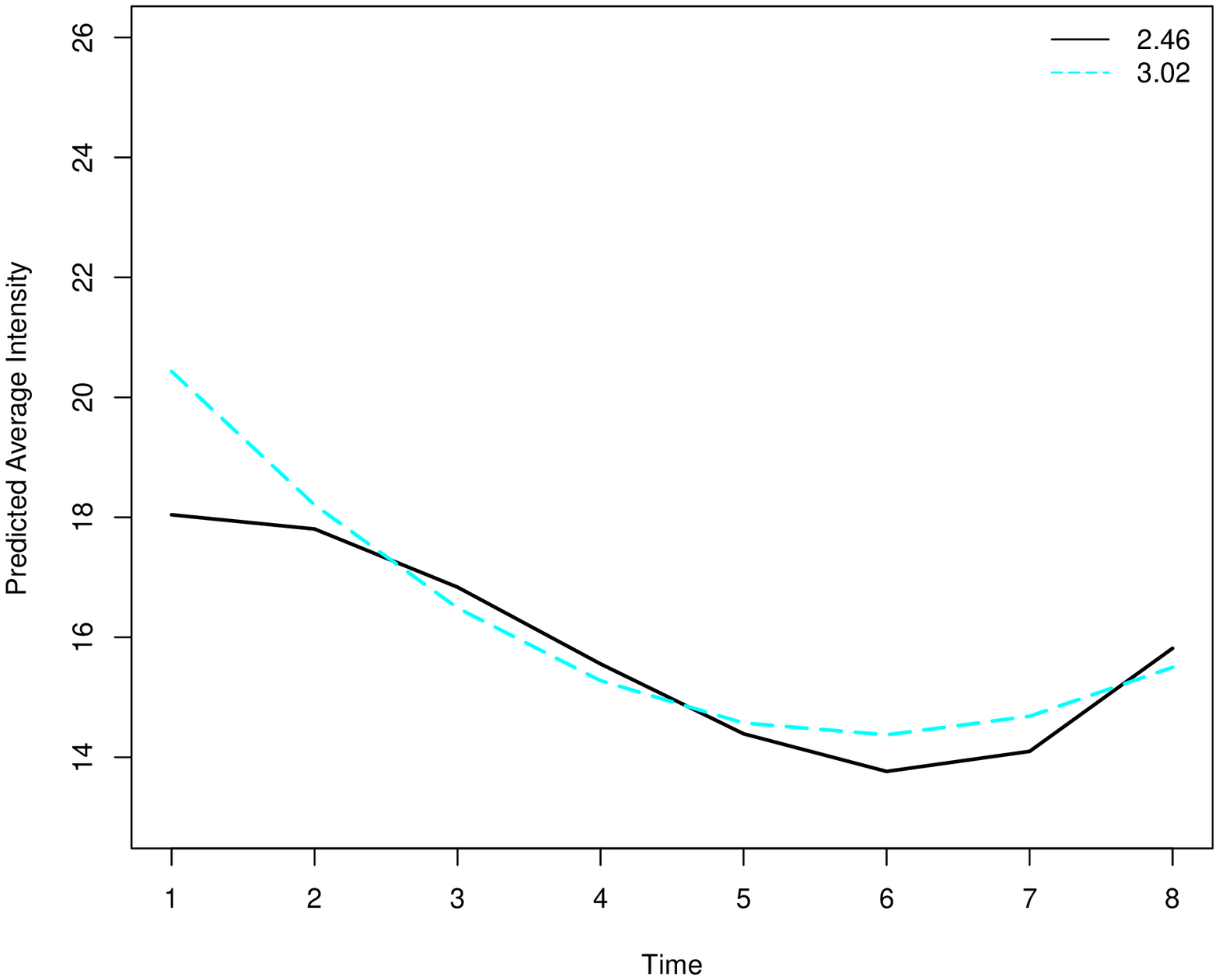}} & 
\subfigure[]{\includegraphics[width=0.45\textwidth,height=0.217\textheight]{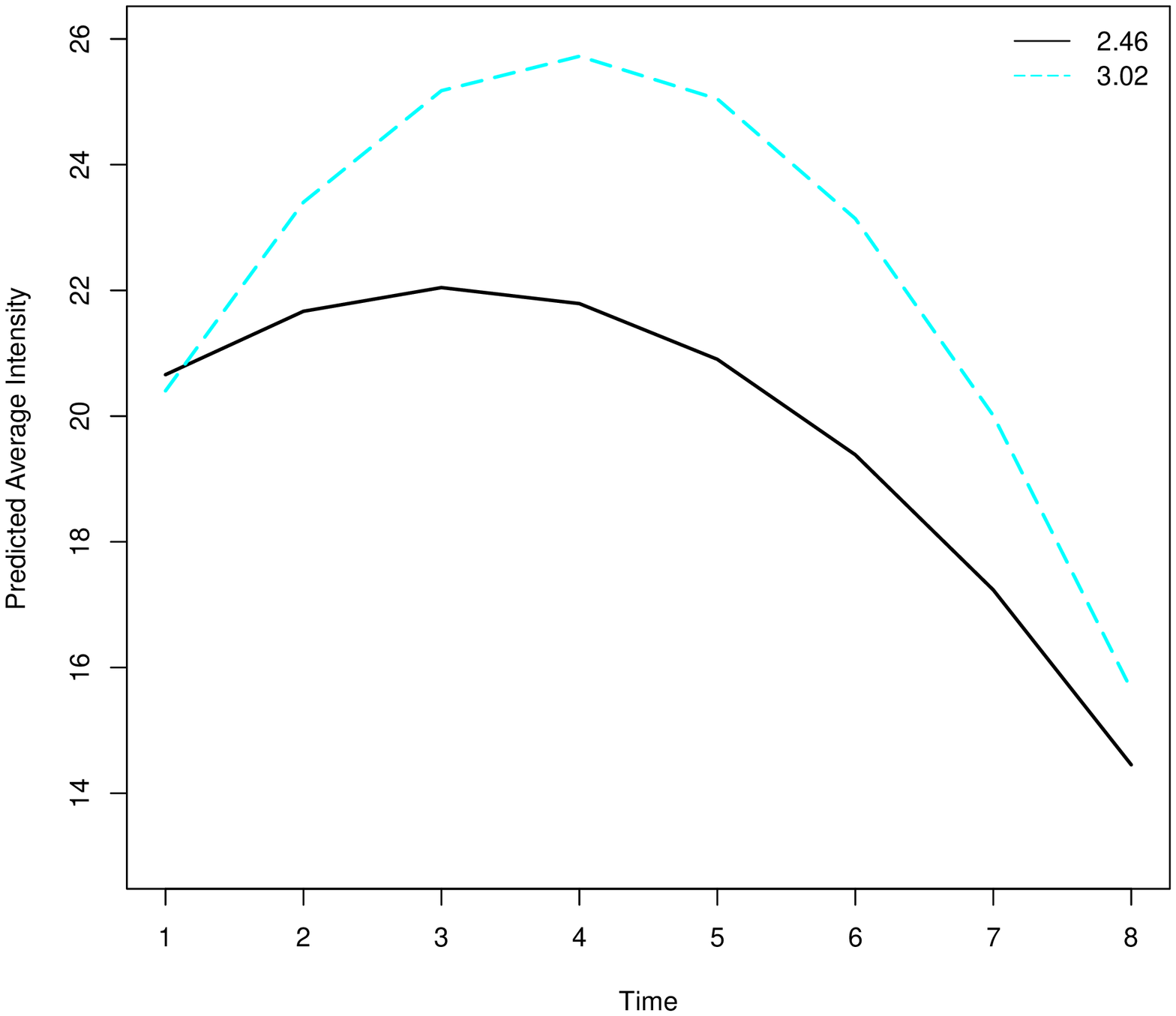} }
\end{tabular}
\caption{\label{fig:2oxy} The LMM predicted average intensities of the two
spectral bins 2.46ppm and 3.02ppm which relate to the metabolite 2-oxoglutarate
in (a) the treatment group and (b) the control group. }
\end{figure}

A total of six spectral bins were highlighted as evolving in the treatment group
and  seven in the control group. There is considerable overlap between the two
sets of evolving bins, with 3.42ppm evolving in the control group only. While
some of the common spectral bins had the same evolution pattern, some differed. 
In particular, the spectral bins 2.46ppm and 3.02ppm relating to the
2-oxoglutarate metabolite were predicted to have opposite quadratic effects in
the treatment group and in the control group. Figure \ref{fig:2oxy}, which shows
the predicted average intensities for these two spectral bins only in both
treatment groups, clearly illustrates this phenomenon. The biological basis of
the diverse response of this metabolite will be investigated in future
metabolomic experiments. \\

\subsection{Assessing model fit}
\label{sec:modelfit}

As with any applied statistical analysis, the modelling assumptions employed need to be assessed to ensure valid inference. In the case of the DPPCA model, the modelling assumptions are the multivariate Gaussian distribution for the latent variables and the error terms, and the stochastic volatility model assumed to control the evolution of the latent variables over time. Posterior predictive model checking \citep{Gelman2003} was employed to assess these modelling assumptions. Replicated data were simulated  from the posterior predictive distribution and compared to the observed data from each treatment group. Given the multivariate nature of the data, the replicated and observed data were compared by examining the mean absolute deviations (MADs) between the covariance matrix of the observed data and the covariance matrix of the replicated data at each time point (\cite{Ansari2002}). The resulting MADs suggested that the DPPCA model fits well since the vast majority of the deviations were close to zero. A 
histogram of the MADs is available in the Supplementary Material. There were some large MADs (6\% of MADs were $>1$ for the treatment group data and 4\% for the control group data) but given the large number of covariance parameters being compared, this was not viewed as sufficient evidence of invalid assumptions and poor model fit.  The few large MADs may arise due to the fact that the number of latent dimensions was fixed at 2 (for visual substantive reasons), and that some parameters were constrained (for reasons of parsimony). Fitting a higher dimensional and less parsimonious model to the time course metabolomic data is an area of further research. 

\section{Discussion}
\label{sec:discussion}

analysing longitudinal data from metabolomics studies is problematic due to the
dimensionality of the data, the correlated metabolites and correlation structure
due to repeated measurements over time. Many currently existing approaches to
analysing such data sets either have the limitation of confounding treatment
variation with variability due to the longitudinal nature of the data or they
ignore the fact that metabolites do not work independently of each other. Here
the DPPCA methodology has been proposed which combines probabilistic PCA and
stochastic volatility models to disentangle the two types of variation in the
data, while also accounting for its high-dimensionality. \\

The DPPCA model successfully addressed the aims of the metabolomic study i.e. visualising the metabolomic trajectories through time, quantifying the effect of time, and  highlighting metabolites which evolve over time. Importantly, the DPPCA model highlighted the contrasting behaviour of the 2-oxoglutarate metabolite between the two treatment groups under study. Future work will examine further this contrasting behaviour. \\

Many areas of further research naturally arise from the DPPCA model. From a practical
viewpoint, fitting the DPPCA model is computationally expensive, mostly
due to the costly sampling of the log volatilities. Several approaches
to sampling log volatilities 
for SV models are suggested and reviewed by \cite{Jacquier94, kim&shephard98}
and \cite{Platanioti05}. Further work in this area would expedite the
convergence of the MCMC chain. Also, while data from 16 times points were collected, only 8 time points were analysed here, due to missing data. Imputation of such data would potentially be feasible within the model fitting algorithm.\\

Motivated by the real application area, only principal subspaces of dimension 2 were considered here; clearly the choice of dimensionality can be viewed as a model selection issue and any of the myriad of approaches to model selection in the Bayesian paradigm by evaluating the
marginal likelihood  could be employed; \cite{Friel12} provide a review of such
approaches. However, it is  anticipated that such approaches would be
computationally expensive in the setting of the DPPCA model. \cite{minka2000} proposes a computationally efficient approach to selecting the optimal dimensionality in PCA, which might also provide a possible solution to the model selection problem here. \\

In terms of the DPPCA model itself, the manner in which the dynamics are modelled in the DPPCA model raises further research questions. Alternative approaches to modelling the time dynamics should be examined, for example (as suggested by a referee) using state-space models for the loadings matrix. Further, research into a random effects PPCA model to model such longitudinal metabolomics data is underway \citep{nyamundanda13}. The DPPCA approach proposed here can be thought of as an approach to identifying the subset of influential variables, which are then analysed via LMMs to highlight those which are time evolving. Hence, the issue of multiple testing is reduced but not eradicated under the DPPCA model; this could be addressed by employing a hierarchical modelling framework \citep{Gelman2003}. Further, the proposed DPPCA approach to highlighting time evolving metabolites requires a two step process: fitting a DPPCA model, followed by fitting LMMs. A more elegant approach would combine the ideas underlying both models into a single model. Clearly the development of the DPPCA model gives rise to many and varied areas of future work.

\bibliographystyle{Chicago}
\bibliography{NyamundandaEtAl}

\end{document}